%% file: ms.tex
\pdfoutput=1
\documentclass[sigconf]{acmart}
\setcopyright{none}
\renewcommand\footnotetextcopyrightpermission[1]{}
\usepackage{setspace}
\usepackage{setspace}
\usepackage{booktabs} % For formal tables
\usepackage{soul,xcolor}
\usepackage{listings}
\usepackage{amsmath}
\usepackage{color}
\usepackage{xcolor}
\usepackage{lipsum}
\usepackage{comment}
\usepackage[title]{appendix}
\usepackage{multirow}
\usepackage{booktabs}
\usepackage{xspace}
\usepackage{graphicx}
\usepackage{caption}
\usepackage{pdfpages}
\usepackage{setspace}
\usepackage{subfiles}
\usepackage{booktabs}
\usepackage{array}
\usepackage{subcaption}
\usepackage[utf8x]{inputenc}
\usepackage[font=small,labelfont=bf]{caption}
\usepackage{url}

\setcopyright{rightsretained}
\usepackage{enumitem}
\setitemize{noitemsep,topsep=0pt,parsep=0pt,partopsep=0pt}
\RequirePackage{filecontents}
\newcolumntype{P}[1]{>{\centering\arraybackslash}p{#1}}

\usepackage{lipsum}
\usepackage{pgfplots,pgfplotstable}
\pgfplotsset{compat=1.4}
\usetikzlibrary{pgfplots.groupplots}

\usepackage{palatino}
\usepackage{tikz}
\usetikzlibrary{shapes, arrows}

\definecolor{RYB1}{RGB}{141, 211, 199}
\definecolor{RYB2}{RGB}{255, 255, 179}
\definecolor{RYB3}{RGB}{190, 186, 218}
\definecolor{RYB4}{RGB}{251, 128, 114}
\definecolor{RYB5}{RGB}{128, 177, 211}
\definecolor{RYB6}{RGB}{253, 180, 98}
\definecolor{RYB7}{RGB}{179, 222, 105}

\pgfplotscreateplotcyclelist{colorbrewer-RYB}{
{RYB1!50!black,fill=RYB1},
{RYB2!50!black,fill=RYB2},
{RYB3!50!black,fill=RYB3},
{RYB4!50!black,fill=RYB4},
{RYB5!50!black,fill=RYB5},
{RYB6!50!black,fill=RYB6},
{RYB7!50!black,fill=RYB7},
}

\pgfplotsset{
    select row/.style={
        x filter/.code={\ifnum\coordindex=#1\else\fi}
    }
}

\newcommand{\name}{Sentinel\xspace}
\newcommand{\squishlist}{
   \begin{list}{$\bullet$}
    { \setlength{\itemsep}{0pt}      \setlength{\parsep}{0pt}
      \setlength{\topsep}{-3pt}       \setlength{\partopsep}{0pt}
      \setlength{\listparindent}{-2pt}
      \setlength{\itemindent}{-5pt}
      \setlength{\leftmargin}{1em} \setlength{\labelwidth}{0em}
      \setlength{\labelsep}{0.5em} }
}

\newcommand{\squishend}{
    \end{list}  }

\definecolor{dong}{RGB}{0,0,0} %{0,0,0}%{255,0,0} 
\definecolor{dong2}{RGB}{0,0,0}%{0,0,0}%{0,0,255}

\begin{document}

\title{\name: Runtime Data Management on Heterogeneous Main Memory Systems for Deep Learning}
%\titlenote{Produces the permission block, and copyright information}
%\subtitle{Extended Abstract}
%\subtitlenote{The full version of the author's guide is available as  \texttt{acmart.pdf} document}
%\begin{comment}
\author{Jie Ren}
\affiliation{
  \institution{University of California, Merced}
}
\email{jren6@ucmerced.edu}

\author{Jiaolin Luo}
\affiliation{
  \institution{University of California, Merced}
}
\email{jluo38@ucmerced.edu}

\author{Kai Wu}
\affiliation{
  \institution{University of California, Merced}
}
\email{kwu42@ucmerced.edu}

\author{Minjia Zhang}
\affiliation{
  \institution{Microsoft Research}
 }
 \email{minjiaz@microsoft.com}
 
\author{Dong Li}
\affiliation{
  \institution{University of California, Merced}
 }
 \email{dli35@ucmerced.edu}
%\end{comment}

\begin{abstract}
Software-managed heterogeneous memory (HM) provides a promising solution to increase memory capacity and cost efficiency. %and avoid limitation of exiting memory technologies. 
However, to release the performance potential of HM, we face a problem of data management. Given an application with various execution phases and each with possibly distinct working sets, we must move data between memory components of HM to optimize performance.  The deep neural network (DNN), as a common workload on data centers, imposes great challenges on data management on HM. This workload often employs a task dataflow execution model,  and is featured with a large amount of small data objects and fine-grained operations (tasks). This execution model imposes challenges on memory profiling and efficient data migration.
%because of the concerns on profiling accuracy and overhead; it also imposes challenges on data migration because of the concerns on page sharing between data objects with different access patterns. 

We present \name, a runtime system that automatically optimizes data migration (i.e., data management) on HM to achieve performance similar to that on the fast memory-only system with a much smaller capacity of fast memory. To achieve this, \name exploits domain knowledge about deep learning to adopt a custom approach for data management. \name leverages workload repeatability to break the dilemma between profiling accuracy and overhead; It enables profiling and data migration at the granularity of data objects (not pages), by controlling memory allocation. This method bridges the semantic gap between operating system and applications. By associating data objects with the DNN topology, \name avoids unnecessary data movement and proactively triggers data movement. \textcolor{dong}{Using only 20\% of peak memory consumption of DNN models as fast memory size, 
%performance of \name is the same or comparable 
\name achieves the same or comparable performance (at most 8\% performance difference) to that of the fast memory-only system on common DNN models; \name also consistently outperforms a state-of-the-art solution by 18\%.} 

\end{abstract}

\maketitle
%\name

\input text/intro
\input text/background
\input text/characterization
\input text/design

\input text/impl
\input text/evaluation

\input text/related_work
\input text/conclusion

%\bibliographystyle{ACM-Reference-Format}
\bibliographystyle{plain}
%\bibliography{sample-bibliography}
\bibliography{kai,li,jie}
%\end{spacing}
%\bibliography{main}
\clearpage

\end{document}

%% file: text/intro.tex
\section{Introduction}

%Emergence of heterogeneous memory
%Hardware heterogeneity, as a practical solution to enable scalable performance and high power efficiency, has been commonly employed by modern computing systems. 
Heterogeneous memory (HM) is an emerging memory architecture, complementary to the existing heterogeneity in processing units (e.g., GPU and FPGA). Within HM, multiple memory components with different technologies are combined to construct main memory within and cross compute nodes. HM brings a promising solution to increase memory capacity, avoid limitation of existing memory technologies, and increase energy efficiency. With the emerging technologies such as non-volatile memory (NVM) and high-bandwidth memory (HBM)~\cite{hbm}, HM is expected to be more common.

%The definition for data management
A typical HM system consists of multiple types of memory components with varying properties (e.g., bandwidth, latency, and capacity). As a result, HM raises a problem of data management and migration. Given an application with various execution phases and each with possibly distinct working sets, we must move data between memory components to optimize performance. Ideally, hot memory pages that are accessed by the running execution phase should be placed in the fastest memory component with the best latency or bandwidth, while other memory pages are filled into other memory components. 

%``The critical operating system support needed to enable the vision of efficiently moving data as programs navigate different phases of execution, each with potentially distinct working sets, is efficient page management and migration. Regardless of configuration, to optimize for performance, ideally the hottest pages will be placed in the fastest memory node (in terms of latency or bandwidth) until that node is full, the next-hottest pages will be filled into the second-fastest node up to its capacity, and so on. Then as programs execute, these pages must be constantly re-organized based on their hotness to retain maximum workload performance.'' 

The data management problem is more complicated, when we consider memory size. In a public cloud, the user is charged not only in terms of processors, but also in terms of memory size. Fast memory in HM tends to be more expensive (e.g., \textcolor{dong}{in the google cloud, regular DDR is 24x more expensive than fast SSD}). The cost of using fast memory is accumulated throughout application execution, making the cloud service less affordable for time-consuming applications. Also, the total cost of ownership (TCO) for fast memory increases quickly (e.g., the average price of DRAM DDR4 (a common fast memory) increased by 2.3x between 2016 and early 2019~\cite{ram_price, ram_price2}), which motivates the data center to reduce the usage of fast memory as much as possible~\cite{Eisenman:2018:RDF:3190508.3190524}. In general, reducing fast memory size without performance loss becomes a critical optimization target to reduce operation cost~\cite{DBLP:journals/corr/abs-1901-10938, Eisenman:2018:RDF:3190508.3190524}.  

%DNN is important
Our work focuses on data management on HM for training the deep neural network (DNN). DNN is a common workload on data centers. %However, training DNN can be resource-demanding and time-consuming. 
Given the success of DNN in many applications and growing trend of training high-quality models specific to individual businesses through automated machine learning, training DNN efficiently is increasingly important to reduce business cost and improve utilization of data centers.

Since modern HM typically serves CPU, we study CPU with HM for DNN training. Using CPU for training is commonly supported by hardware vendor (e.g., Intel MKL-DNN~\cite{intelMKL} and ARM Compute Library~\cite{arm_cl}), and has the following four benefits. 
\textit{First}, the trend of democratizing DNN~\cite{democratization_ai} makes CPU an appealing solution. compared with GPU, CPU is more approachable and affordable, especially for personal users or small-sized enterprises. \textit{Second}, some data centers do not have GPU and simply use CPU for training. Such examples include the Cori~\cite{lbnl_cori} at Lawrence Berkeley National Lab and Stampede2~\cite{tacc_stampede} at TACC for scientific machine learning~\cite{ 8658402, Kurth:2018:EDL:3291656.3291724, Mathuriya:2018:CUD:3291656.3291743,  tacc_ml_CPU_training}. \textit{Third}, for those DNN models that lack thread level parallelism, GPU can perform worse than CPU. For example, training some CNN (e.g., the wide-and-deep model~\cite{DBLP:journals/corr/ChengKHSCAACCIA16}) and some deep reinforcement learning models (e.g., DQN~\cite{Hasselt:2016:DRL:3016100.3016191}), CPU performs faster than GPU. In our evaluation, using 8-core Intel i7-7700K CPU and NVIDIA Titan XP GPU to train the wide-and-deep model, the training throughput on CPU is 4x of on GPU (763 and 196 global steps per second for CPU and GPU respectively). \textit{Fourth}, on a public cloud, CPU is cheaper than GPU. For example, on the google cloud, one vCPU %core of the default Intel Xeon E5 (v4) CPU 
is only 1/46 and 1/78 of NIVIDA P100 and V100 GPU, in terms of cost per hour. When the training throughputs on CPU and GPU are comparable, using CPU can be easily more cost effective. 
%\textit{Fourth}, the CPU-based machine provides a much larger memory capacity than the GPU one, allowing more use cases such as co-locating multiple DNN training workloads in a virtualization environment or automatic machine learning~\cite{google_auto_ml, DBLP:journals/corr/BailisORZ17, aws_auto_ml}.  

%challenges.
The workload characterization of training DNN imposes unique challenges on data management on HM. \textit{First}, training DNN is typically featured with a large amount of data objects smaller than a memory page (4KB). Profiling memory accesses to those data objects to make the decision of data placement on HM is challenging, because those small data objects can share memory pages, creating difficulty to track memory accesses for individual data objects. 

\textit{Second}, training DNN often employs a task dataflow execution model, where the whole computation is decomposed into a large amount (thousands or even millions) of operations in a single training step. Those operations (e.g., matrix multiplication and 2D convolution) represent various execution phases with diverse memory accesses patterns. Many of those operations can run in parallel and take short execution time. Placing data objects on HM for those operations has high requirements on the promptness of making the data placement decision and data migration, in order to make best use of fast memory and enable high performance. 

\textit{Third}, the semantic gap between operating system (OS) and application (DNN in our case) can make data migration less efficient. From the OS point of view, the memory page is the primitive granularity for memory profiling and data migration. However, from the application point of view, the data object (or the tensor in the language of DNN) is the primitive granularity for operations to access and compute. Ideally, we want to migrate data at the granularity of data objects to enable high performance of operations. However, this is conflicting with the abstract (i.e., the memory page) OS uses to manage data. 

%The existing solutions cannot work. 
Unfortunately, the existing data management methods cannot address the above challenges well. Many methods~\cite{Thermostat:asplos17,RAMinate:socc16,heteros:isca17,sc18:wu,unimem:sc17} use a sampling-based approach to profile memory pages to \textcolor{dong2}{avoid expensive profiling overhead}, which can miss memory accesses for small data objects and lead to incorrect data migration decision; Furthermore, the application semantics is often missed, leaving many opportunities to improve performance on the table.

In this paper, we present \textit{\name}, a runtime system that automatically optimizes data migration (i.e., data management) on HM to achieve performance similar to that on the fast memory-only system with a smaller size of fast memory. To achieve this, \name exploits domain knowledge about deep learning to adopt a custom approach for data management. %exploits the unique characteristics of the DNN workload. 

% A brief overview of our solution;
\name leverages the fact that a DNN training workload has high repeatability and hence highly predictable. Such a workload comprises of millions of training steps, each of which goes through the exactly same computation graph and operations. As a result, \name uses a few training steps for profile measurements. In addition, \name enables profiling and data migration at the granularity of data objects (not pages), by controlling memory allocation. This method bridges the semantic gap between OS and application. More importantly, it allows the runtime system to employ the limited domain information for runtime data management: By associating data objects with the DNN network topology, \name avoids unnecessary data movement and proactively triggers data movement.

%``a DNN training job is a unique workload from a system perspective. Each job comprises of millions of mini-batches, where each mini-batch looks at a small number of training inputs. Because each mini-batch typically goes through exactly the same computation graph, mini-batches are identical to each other from a resource-usage and performance viewpoint.''

%\textbf{``Third, we use the initial profile measurements during the online exploration as signals to prune the dynamic state space in an intelligent manner.''}

Data migration faces a fundamental tradeoff between migration frequency and performance benefit. To save the capacity of fast memory, we want to frequently move data between fast and slow memory, such that we can timely move unused data out of fast memory and move to-be-used data into it. However, frequent data movement can be exposed to the critical path and cause performance loss. Hence, choosing an appropriate migration interval is critical to reduce memory capacity and avoid performance loss. Guided by the profiling results, \name explores the optimal migration interval for best performance.
%and theoretically guarantees the optimal one.

% summary of contributions
The key contributions of our work are as follows.
\begin{itemize}
    \item \textbf{Performance characterization.} We use a data object-centric (instead of page-centric) approach to systematically analyze the performance of DNN, a typical task dataflow workload. We identify and leverage the unique characteristics of such a workload to direct data management at runtime on HM. 
    
    %\textbf{Profiling method and results}, ``We explore a simple end-to-end heterogeneous memory profiling and placement policy. Unlike existing implementations and other recent proposals, our system does not swap data out to disk, nor does it make portions of memory unavailable to profile accesses to them via page faults. Instead, it simply repurposes the existing OS active/inactive page lists. Therefore, our approach imposes no profiling overhead on systems which may not need its functionality''. 
    
    %``We identify and leverage the unique characteristics of the deep learning workload to do extreme tailoring or custom-wiring of the infrastructure for a specific job, resulting in significant efficiency gains;''

\item \textbf{Runtime system.} We propose and evaluate a runtime system for optimizing data placement and migration; We determine the optimal migration interval based on theoretical analysis, dynamic profiling, and DNN domain knowledge. %to minimize the fast memory size while completely hiding data migration cost. 

%%We propose and evaluate a new architectural frameworkfor optimization of such custom workloads, with aggres-sive multi-version compilation at a whole-program level that performs parallel exploration of independent choices by using fine-grained profiling.

%We developspg-CNN, an op-timization  framework  for  CNNs,  which  introduces  andintegrates  three  key  components,  i)  an  alternate  sched-ule  using  GEMM-in-Parallel,  improving  scalability,  ii)  astencil-based  code  generator,  improving  per-core  perfor-mance, and iii) a sparse code generator, exploiting sparsityand improving goodput, into a unified code generator thatproduces optimized codes for various CNN computations(Section 4).
%\textbf{Framework. fundamental tradeoff. How to determine migration interval: migrate too often (overhead); migrate not too often (make best use of fast memory). Migration interval too short (migration cannot finish before the execution happens); migration interval too large (too many short-lived variable and no DRAM space). }

\item \textbf{Evaluation.} We evaluate \name using TensorFlow. The evaluation results show that using only 20\% of peak memory consumption of DNN models as the fast memory size, Sentinel achieves the same or comparable performance (at most 8\% performance difference) to that of the fast memory-only system on common DNN models. \name also consistently outperforms a state-of-the-art solution by 18\%.

%Compared with a state-of-the-art runtime data management method~\cite{Yan:ASPLOS19}, \name performs 18\% better on average.} 
%performance  improvement  of  up  to  \textcolor{red}{xxx}  on  real-world benchmarks, compared with the state-of-the-art runtime data management method~\cite{Yan:ASPLOS19}. \name provides performance close to that of the fast-memory only system, while significantly saves the fast memory size by \textcolor{red}{xxx}, compared with xxx. 
 
\end{itemize}

%%%%%%%%%%%%%%%%%%%%%%%%%%%%%%%%%%%%%%%%%%%%%%%%

\begin{comment}
CPU for machine learning. 

We focus on CPU as the training platform because of following reasons:
(1) The widespread deployment of CPUs makes this hardware platform an attractive target for machine learning model training. 
(2) The optimizations on GPU lack of generality. 
(3) In terms of programmability, porting the original
code to GPU kernels requires significant programming efforts; 
(4) GPU has strong bias towards certain application.
Existing effort has been made in academic and industry to optimize machine learning training on widely available CPU based architecture. Adam~\cite{Chilimbi:2014:PAB:2685048.2685094} from Microsoft. Intel MKL-DNN\cite{intelMKL} has been proposed to accelerate machine learning training. 

\end{comment}

%% file: text/background.tex
\section{Background}
\label{sec:bg}

We provide a brief background on the training process of DNN models and HM. 
%and the current state of using heterogeneous main memory systems.

\subsection{Training Deep Learning Models}
\label{sec:bg_training}
%\textcolor{red}{introduce the basic concepts of model training, such as layers and training accuracy; See the astra paper (ASPLOS paper)}
%\textcolor{red}{introduce the task dataflow model (execution model)}
%\textcolor{red}{discuss that the DL models are highly predictable. See the astra paper.}
%\textcolor{red}{discuss that the CPU-based training makes sense.}

A typical DNN model comprises of a stack of \textit{layers} each of which is a group of neurons. Each neuron in a layer computes a non-linear function of the outputs of neurons in the preceding layer, using a set of weights. Training DNN often involves a large number of iterative training steps. In each step, a batch of training samples are fed into DNN. Performance of each step (e.g., execution time and memory access pattern) remains stable across steps~\cite{liu:micro18,ipdps19_liu,DBLP:conf/asplos/SivathanuCSZ19}. The above characteristics allow us to use dynamic profiling of the first few training steps to improve performance of the following steps. 
%\textcolor{red}{(add a discussion on dynamic graph here?)}

Training DNN often uses a machine learning framework, such as TensorFlow~\cite{tensorflow2015-whitepaper}, PyTorch~\cite{pytorch}, and MXNet~\cite{mxnet}. These frameworks use a dataflow execution model where the whole workload of DNN is modeled as a directed graph composed of a set of nodes. \textit{Operations}, such as 2D convolution, matrix multiplication, and array concatenation, are implemented by the frameworks as primitives. Those operations are represented as nodes in the dataflow graph. Within the graph, edges between nodes capture dependencies between nodes. %An operation is ready to run, as long as its dependencies (control or data dependencies) are resolved. 

\subsection{Data Management on Heterogeneous Memory}
%%Software-controlled HM is emerging. Examples of this system include non-volatile 3D XPoint memory alongside DRAM, high-bandwidth multi-channel DRAM alongside DDR4 in Intel Knight's Landing, distributed memory components in the disaggregated memory system. 
The online data management on HM typically involves three fundamental steps: (1) memory profiling, (2) decision making for data migration, and (3) data migration. 

The memory profiling step collects memory access information for pages or data objects; The decision making step uses performance models or caching algorithms to decide which pages to migration for best performance; The data migration step triggers data migration with the goal of reducing data migration overhead.

\textbf{Recent research efforts.}
%Given the promising of HM, data management on HM remains an active area of research. 
The three steps create major optimization targets in the existing research efforts. We list the targets as follows. Our work shares the same optimization targets as the existing efforts.
%\squishlist
\begin{itemize}
    \item Memory profiling must have ignorable impact on application performance while being accurate;
    \item The decision-making process must timely capture those hot data for migration without violating the capacity of fast memory;
    \item The data migration cannot impact  performance. %and/or not exposed to the critical path.  
\end{itemize}
%\squishend

%\textcolor{dong}{To complete the following paragraph, see Section 2.2 in the Nimble paper.}
The existing efforts explore hardware techniques for profiling and facilitate data movement among memory devices~\cite{asplos15:agarwal,hetero_mem_arch,qureshi_micro09, ibm_isca09, Ramos:ics11,gpu_pcm_pact13,hpdc16:wu,row_buffer_pcm_iccd12}. Further studies on software-based techniques use sampling-based approaches for memory profiling to reduce profiling overhead~\cite{Thermostat:asplos17,RAMinate:socc16,heteros:isca17, unimem:sc17, sc18:wu}; They commonly use a caching algorithm, such as the multi-queue~\cite{RAMinate:socc16,Ramos:ics11, 5260554}, FIFO~\cite{Yan:ASPLOS19}, or LRU~\cite{heteros:isca17}. \textcolor{dong}{However, They often trade memory profiling accuracy for low profiling overhead, and hence can lose tracking for small data objects. Also, the process of detecting hot pages may not timely trigger data migration.} 
%They also try to optimize the data migration mechanism using parallel migration, two-sided migration~\cite{Yan:ASPLOS19} or proactive migration~\cite{unimem:sc17}.  
%In addition, the modern OS uses autoNUMA~\cite{autonuma} or the active list~\cite{Yan:ASPLOS19} for data management on HM. 
%\textcolor{dong}{However, autoNUMA offlines pages for profiling, and cannot swap pages between fast and slow memories when there is no space in fast memory. Hence it cannot meet the need of online DNN training; The active list is slow to capture memory access patterns, and cannot effectively direct data placement for small data objects.}

We study a fundamentally new method for data management on HM. %ignore application knowledge which can be used to achieve the optimization targets with less performance overhead. 
Inspired by the trend of using domain specific knowledge for hardware (e.g., AI accelerators~\cite{7738524, Jouppi:2017:IPA:3079856.3080246, 8192478, Shafiee:2016:ICN:3001136.3001139} and Anton for molecular dynamics simulation~\cite{Shaw:2008:ASM:1364782.1364802}) and software (e.g., domain specific language Halide~\cite{halide} and Liszt~\cite{DeVito:2011:LDS:2063384.2063396}), we propose to use the domain knowledge of DNN to direct data placement. This method provides lightweight profiling, accurately captures data hotness, timely trigger data migration, and effectively hides data migration overhead. 
%avoids repeated profiling (hence reducing profiling overhead), accurately captures data hotness, and effectively hide the data migration overhead. 

%``To ensure optimal performance, application developers will rely on both initial page placement policies and follow-up page migration policies to ensure that hot data pages remain within the highest bandwidth or lowest latency memory node during an application’s runtime''

%%%%%%%%%%%%%%%%%%%%%%%%%%%%%%%%%%%%%%%%%%%%%%%%%%%%%%%
\begin{comment}
For instance, multi-GPU systems have been explored
when the batch and model parameters do not fit in the memory of a single GPU. The solution has been to reduce the batch size and make it fit in memory. But this changes the batch size and alters the convergence
\end{comment}

%% file: text/characterization.tex
\section{Analysis and Characterization of Memory Accesses in DNN}
\label{sec:char}
We analyze and characterize memory accesses in DNN and use the analysis results to drive our design.

\subsection{Profiling Framework}
\label{sec:profiling_framework}
%The major profiling challenge is to semantic difference between OS (page) and app (tensors)

%\textcolor{green}{paragraph: How OS level page access information are captured. TLB miss, PTE(page table entry) walk. }
We build a profiling framework for our study.
The profiling framework collects the following information: the number of main memory accesses per data object (tensor), data object size and lifetime.

To collect the above information, the profiling framework includes the support at \textit{both} OS and application levels. At the OS level, \name collects the number of memory accesses at the page level. This is implemented by a software-only solution. In particular, when a page is tracked for access counting, \name sets a reserved bit (bit 51) in its PTE (i.e., poisoning PTE) and then flush the PTE from TLB. When the page is accessed, a TLB miss occurs and triggers a protection fault. \name uses a customized fault handler to count this page access, poison the PTE and flush it from TLB again to track next page access. Poisoning PET only happens during the profiling. After it, poisoning PTE and flushing TLB do not happen. 
%reserving the PTE bit and flushing PTE do not happen.

%\textcolor{red}{poisoning PTEs at the page table(i.e., setting PTE reserved bit) and then flushing the PTE from TLB. When the page be accessed again, a TLB miss occurs and triggers a protection fault. \name implement customized TLB miss handler to record page access information and poison this PTE again to keep recording the next of this page access. Note that \name profiling triggers a customized system call to start collect the number of memory access at page level. When the profiling finishes, another system call will be triggered and PTEs will no longer be poisoned. }

To bridge the semantic gap between OS and application, \textcolor{dong2}{each memory page has only one data object (but a data object can use more than one pages).} %\name allocates no more than one data object in a memory page for profiling, 
Using this method, page-level profiling becomes data object-level profiling. Such memory allocation does not change memory access patterns captured by the hardware caching mechanism in the cache hierarchy, hence providing reliable estimation on memory accesses in main memory. Such memory allocation increases memory footprint. But it happens during the profiling phase of \name on slow memory. After the profiling phase, data objects are re-organized to reduce memory footprint and improve performance. \textcolor{dong2}{Data reorganization happens during memory allocation (see Section~\ref{sec:dyn_profiling_data_org}), and hence does not stop the training process and does not impact performance.} Also, the profiling method does not increase the consumption of fast memory.

At the application level, \name leverages memory allocation and deallocation to get the size and lifetime of data objects.
\textcolor{dong}{Furthermore, \name introduces API that allows the user to annotate DNN to indicate the end of each layer in DNN.}
%Furthermore, \name introduces two APIs that allow the user to annotate DNN to indicate the start and end of each layer in DNN. 
Based on the above infrastructure, \name is able to associate a data object with the DNN model topology of DNN (i.e., we can know which layer(s) a data object is alive). Setting up the association is helpful to direct data migration (Section~\ref{sec:adaptive_dm}).  

Our profiling method uses only one %\textcolor{red}{one}
training step for profiling. During the profiling, \name captures each page read and write by repeatedly poisoning the page. This is expensive because of system calls and TLB misses. However, it does not lose profiling accuracy. Also, considering that a typical DNN training involves millions of training steps, the profiling overhead is easily amortized.  

The traditional profiling methods face a fundamental dilemma between profiling overhead and accuracy. In particular, frequently collecting memory access information brings high profiling accuracy at the cost of large runtime overhead, and vice versa~\cite{Thermostat:asplos17,RAMinate:socc16,heteros:isca17, sc18:wu, unimem:sc17}. Leveraging the repetitiveness of DNN training, \name breaks the dilemma, and enables both high profiling accuracy and low profiling overhead. 

%\textcolor{green}{paragraph: Existing work use OS level profile use sampling based approach intercept TLB miss to capture hot and code pages.}

%\textcolor{green}{paragraph: TensorFlow provides profiling API to profile the input and output tensor for operations. The information is course-gained and not suitable for migration: (1) lack of tensor access info which has been used instead of operations. (not work for short-lived tensor). (2) lack of tensor location information. tensorflow profiling API only provide the information of the pointer points to tensor instead of the real address of tensors. The pointer information is not suitable for page migration. }

%\textcolor{green}{paragraph: app level profiling: profiling tensor access in app runtime ask for a lot of engineering efforts. need to intercept every tensor access in every operation. Not practical, not general.}

%\textcolor{green}{paragraph: Our profiling method. We adopt the combination of profiling in both OS level and app level. In app level we intercept tensor allocator/deallocator to get tensor size and lifetime information; In OS level we poison PTE to trigger protection fault and record every page access.}

%\textcolor{green}{paragraph: Due to the predicative of ML model, profiling one training step without sampling can have more accurate result. Profiling happens on slow memory.}

\subsection{Profiling Results and Analysis}
\label{sec:profiling_results}

\begin{figure}[!t]
\centering
%\vspace{-10pt}
%\includegraphics[width=0.48\textwidth]{ASPLOS2020_Jie/figures/f1_3edit.pdf}
\includegraphics[width=0.48\textwidth]{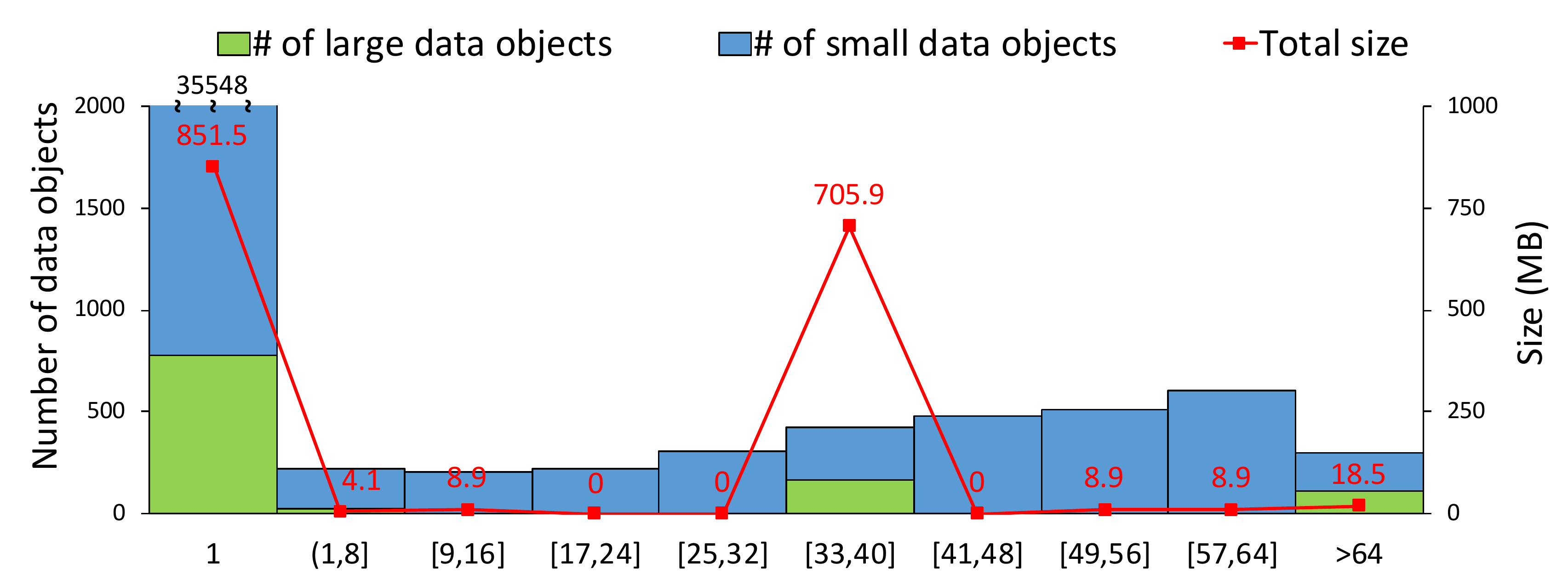}
%\vspace{-10pt}
%\caption{\name tensor access histogram.  x-axis shows tensor access frequency and y-axis shows the number of tensors.}
\caption{Distribution of lifetime of data objects and their sizes for ResNet\_v1-32. Each small data object is smaller than 4KB; Each large data object is no smaller than 4KB. ``>64'' means that the data object survives more than one forward and backward pass (one training step).}
%\textcolor{red}{need to conform}}
%\vspace{-10pt}
\label{fig:fig_lifetime}
\end{figure}
We use the profiling framework to study data objects and their access patterns in DNN. \textcolor{dong2}{We report profiling results for one training step in this section.}

Figure~\ref{fig:fig_lifetime} shows the distribution of lifetime of data objects and their accumulated sizes for ResNet\_v1-32 (the configuration of training is in Table~\ref{tab:models}). ResNet\_v1-32 has 64 layers (in a forward and backward pass).  %(\textcolor{dong}{each residual block has two layers})}.  
%and \textcolor{orange}{consumes 6GB memory. 
\textcolor{dong2}{A data object is alive after it is allocated and before it is freed. The lifetime of a data object is defined in terms of number of layers where the data object is alive.}
%The lifetime of a data object is defined from memory allocation to memory free. The lifetime is shown in terms of layers in Figure~\ref{fig:fig_lifetime}. 
\textcolor{dong2}{Figure~\ref{fig:fig_lifetime} shows that 92\% of data objects have lifetime no longer than one layer. Among those short-lived data objects, 98\% of them is small data objects (smaller than 4KB). }

\textbf{Observation 1}: There are a large number of small data objects with short lifetime in DNN workloads.

In the rest of the paper, we define short-lived data objects as those with lifetime no longer than one layer.

%\textcolor{dong2}{The size of memory space for short-lived data objects is small, and typically bounded by a few GB.} 

\begin{figure}[!t]
\centering
%\vspace{-5pt}
%\includegraphics[width=0.48\textwidth]{figures/tensor_access.pdf}
\includegraphics[width=0.48\textwidth]{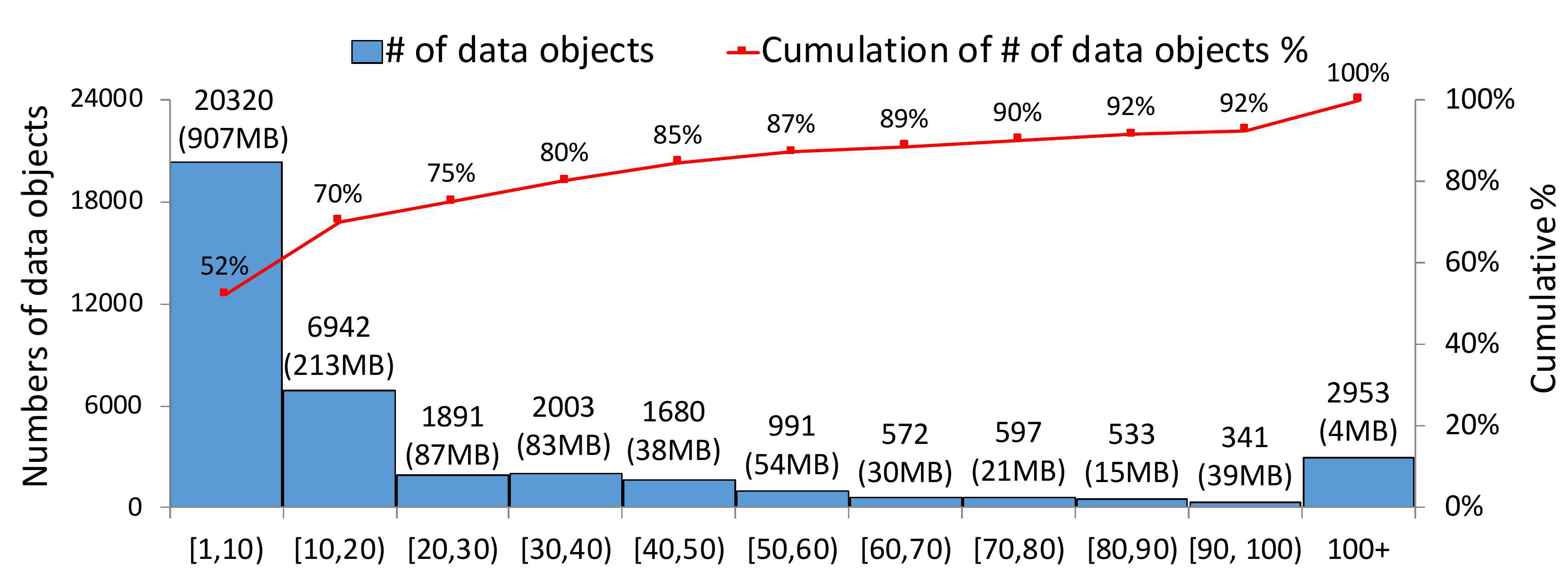}
%\vspace{-20pt}
%\caption{\name tensor access histogram.  x-axis shows tensor access frequency and y-axis shows the number of tensors.}
\caption{Distribution of the number of main memory accesses at the data object level.}
%\vspace{-5pt}
\label{fig:tensor_access}
\end{figure}

\begin{figure}[!t]
\centering
%\vspace{-20pt}
%\includegraphics[width=0.48\textwidth]{figures/small_tensor_access.pdf}
\includegraphics[width=0.48\textwidth]{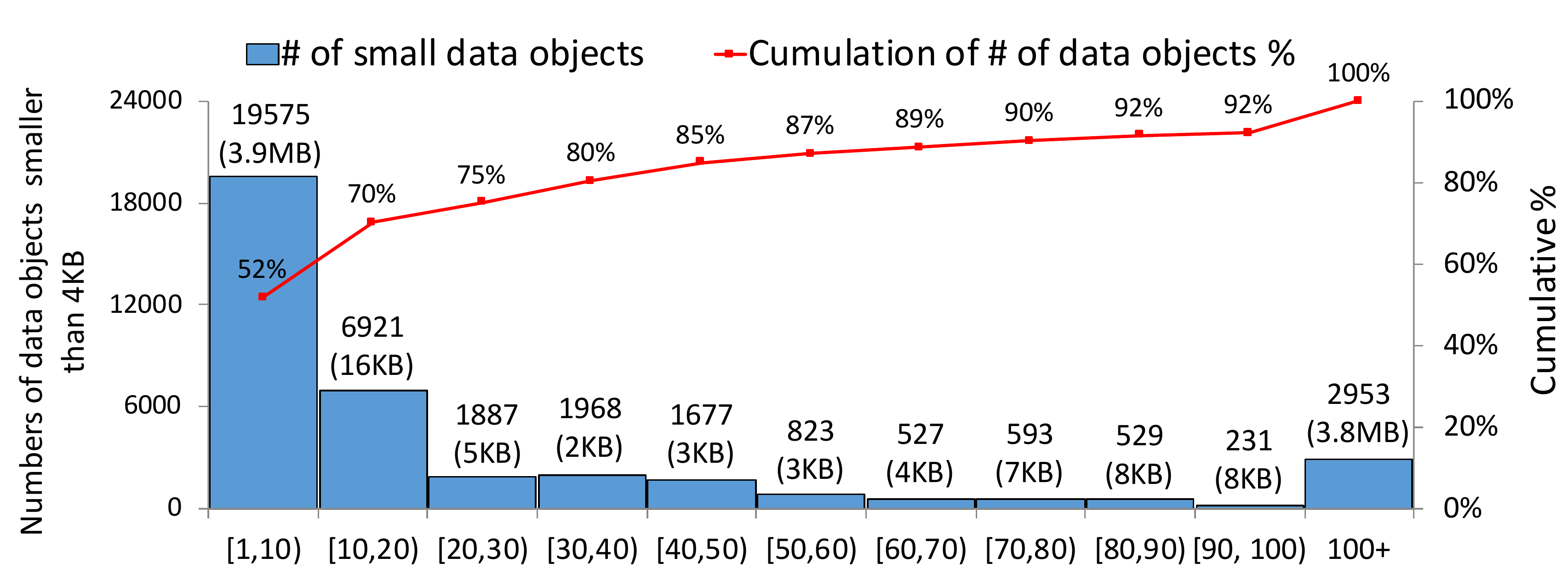}
%\vspace{-25pt}
%\caption{\name tensor access histogram.  x-axis shows tensor access frequency and y-axis shows the number of tensors.}
\caption{Distribution of the number of main memory accesses at the data object level for small data objects (each is smaller than 4KB).}
%\vspace{-10pt}
\label{fig:small_tensor_access}
\end{figure}

%\textcolor{red}{Figure: show the distribution of number of memory accesses to smaller tensors (less than one mem page). Show how many pages those tensors use. Purpose of this figure: (1) page-level false sharing exists (define false-sharing); (2) there is opp to reduce the fast mem size}

Figure~\ref{fig:tensor_access} shows the distribution of the number of main memory accesses at the data object level. 
%We collect the data in the figure after data re-organization. 
%%%\textcolor{green}{We collect the data in figure by allocating each memory page has only one data object.}
The figure shows that a large number of data objects (52.3\% of data objects, using 907 MB, which is 54\% of total memory pages) are accessed less than 10 times. Among them, 98\% of them are small (\textcolor{dong2}{less than 4KB}) and use only 3.9 MB in total, shown in Figure~\ref{fig:small_tensor_access}. On the other hand, some data objects are frequently accessed (having >100 accesses), taking only 4 MB (0.2\% of total memory pages). They are the candidates to be placed into fast memory, and their size is a small portion of total memory pages.

\begin{comment}
\begin{figure}[!t]
\centering
%\vspace{-20pt}
\includegraphics[width=0.48\textwidth]{figures/tensor_access.pdf}
%\vspace{-20pt}
%\caption{\name tensor access histogram.  x-axis shows tensor access frequency and y-axis shows the number of tensors.}
\caption{Distribution of the number of main memory accesses at the data object level.}
%\vspace{-5pt}
\label{fig:tensor_access}
\end{figure}

\begin{figure}[!t]
\centering
%\vspace{-20pt}
\includegraphics[width=0.48\textwidth]{figures/small_tensor_access.pdf}
%\vspace{-25pt}
%\caption{\name tensor access histogram.  x-axis shows tensor access frequency and y-axis shows the number of tensors.}
\caption{Distribution of the number of main memory accesses at the data object level for small data objects (each is smaller than 4KB).}
%\vspace{-10pt}
\label{fig:small_tensor_access}
\end{figure}
\end{comment}
\textbf{Observation 2}: The uneven distribution of hot and cold data objects in DNN provides opportunities for data management. 

\begin{table}[!tbh]
%\vspace{-10pt}
%\small
\centering
\caption{Memory consumption (in one training step) in the original execution and using ``one data object per page'' in the profiling step. ``prof.'' stands for ``profiling''.} 
\label{tab:mem_consume}
\begin{tabular}{|c|c|c|}
\hline
memory consumption                                                       & in prof. & Orig. exe.  \\ \hline
all data objects                                                         & 1.97 GB   & 1.57 GB    \\ \hline
\begin{tabular}[c]{@{}c@{}}data objects \\ smaller than 4KB\end{tabular} & 152 MB    & 0.45 MB    \\ \hline
\end{tabular}
\end{table}

%\textcolor{red}{Figure: show the distribution of number of memory accesses at page level. Purpose of this figure: show the difference between page-level and tensor-level profiling.}

\textcolor{dong2}{Table~\ref{tab:mem_consume} shows memory consumption for two cases: (1) the original execution and (2) using ``one data object per page'' in the profiling step. In the original execution, small data objects takes only 0.45MB, but using one data object per page, they take 152 MB.} This indicates that small data objects commonly share pages with other data objects. 

\begin{figure}[!th]
\centering
%\vspace{-20pt}
%\includegraphics[width=0.48\textwidth]{figures/mem_size.pdf}
\includegraphics[width=0.48\textwidth]{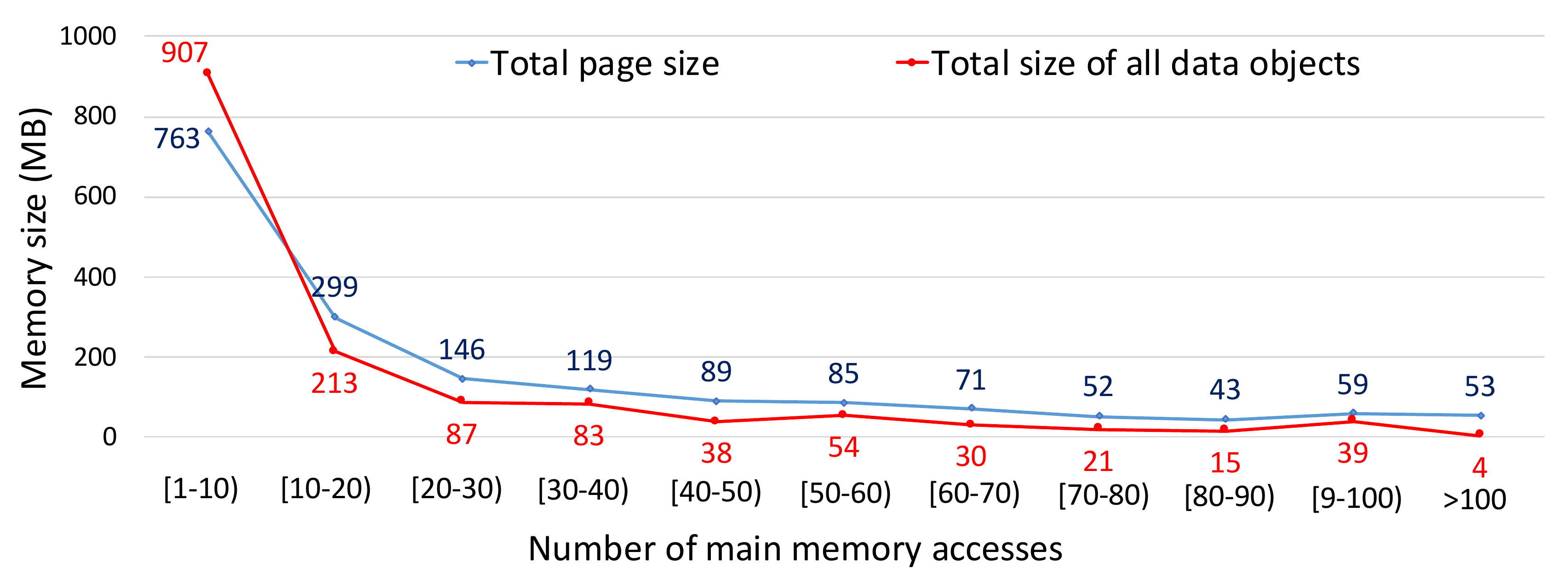}
%\vspace{-25pt}
%\caption{\name tensor access histogram.  x-axis shows tensor access frequency and y-axis shows the number of tensors.}
\caption{Distribution of the number of main memory accesses at the levels of pages (in the original execution), data objects, and small data objects.}
%\vspace{-10pt}
\label{fig:size_diff}
\end{figure}

Figure~\ref{fig:size_diff} shows the distribution of the number of main memory accesses at different levels, including at the data object level already shown in Figures~\ref{fig:tensor_access} and~\ref{fig:small_tensor_access}, and page level in the original execution. The figure shows that for less frequently accessed data objects (having 1-10 accesses), the total size of data objects (907 MB) is larger than the total page size (763 MB) in \textcolor{dong2}{the original execution}. 

This result is interesting, because if alive data objects fall into the same pages, the size of the data objects should be smaller than or equal to the size of pages.  
%\textcolor{green}{if data objects with similar access time falling into the same pages, the size of the data objects should be smaller than or equal to the size of pages. }
Our result is against the above rationale, which suggests that some data objects actually do not fall into those 763MB-pages \textcolor{dong2}{in the original execution}. This means \textcolor{dong2}{in the original execution}, those data objects fall into other pages that are counted as more frequently accessed. In other words, those data objects share the pages with other data objects that may have different preference for data placement. We refer to the above result as \textit{page-level false sharing} in the rest of the paper.

%This indicates that some of these data objects are placed into the same page. Since the data objects in the same page can be accessed at different execution time, such a memory page can be . An example of this case is xxxx. We call this case page-level false sharing in the rest of the paper.

%%We notice that 48\% of pages is not frequently accessed (less than ten times). %Those pages have great potential to be placed in slow memory in most of the training time. Those frequently accessed pages (having >500 accesses) only take about \textcolor{red}{xxx} MB (3.3\% of total memory pages). This result is different from Figure~\ref{fig:tensor_access} at the tensor level. page-level false sharing.
%Those hot pages are candidates to be placed into fast memory. The small size of those hot pages shows great potential to use a small fast memory.

\begin{comment}
\begin{figure}
\centering
%\vspace{-20pt}
\includegraphics[width=0.48\textwidth]{figures/page_access.pdf}
%\vspace{-25pt}
\caption{Distribution of the number of main memory accesses at the page level without data re-organization.}
%\caption{x-axis shows the page access frequency and the y-axis shows the number of pages.}
%\vspace{-10pt}
\label{fig:page_access}
\end{figure}
\end{comment}

\textbf{Observation 3}: Page-level false sharing exists in DNN. The page-level profiling (not data object-level) for data management can be misleading because of page-level false sharing. 

%\textcolor{red}{Figure: show the number of tensors and their sizes per layer. Purpose of this figure: to show the variance of tensor usages across layers.}

%\textcolor{green}{Figure~\ref{fig:page_access} shows main memory page access distribution in page level. More than 50\% of pages access time is less 10 times. Figure~\ref{fig:size_diff} shows the accumulated size with different profiling method. Page-level profiling can over-estimate data object access because of false-sharing.}

%% file: text/design.tex
\section{Design}
\subsection{Overview}
\begin{figure}[t!]
	\centering
	\includegraphics[height=0.11\textheight]{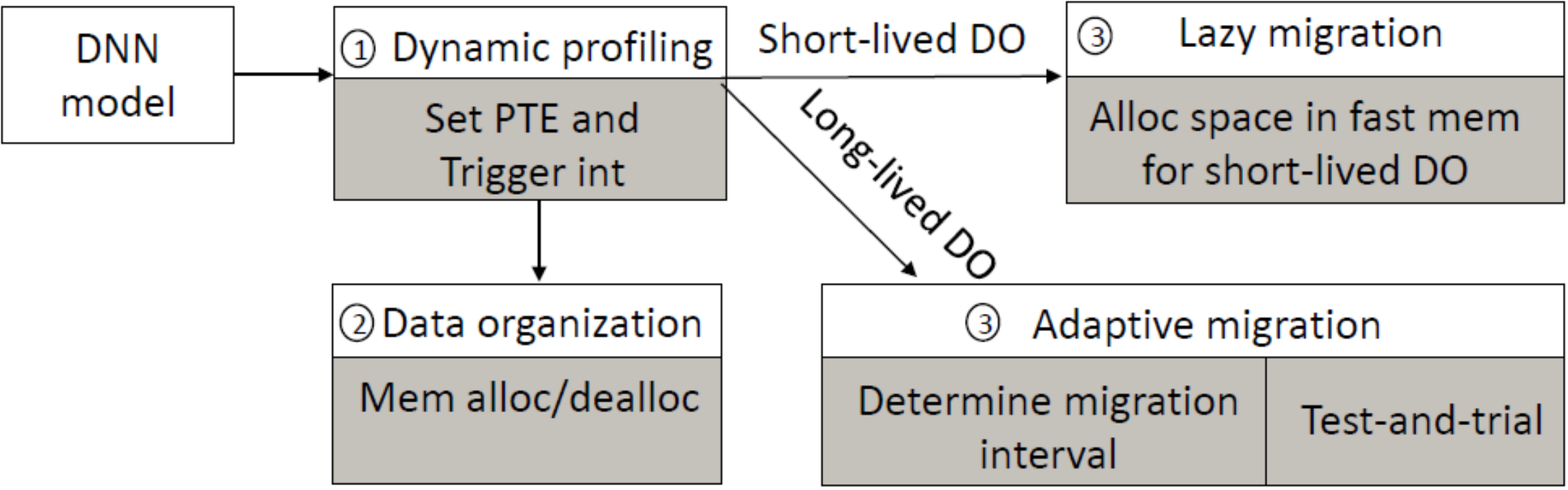}
	\caption{Overview of \name. ``DO'' stands for ``data object''.  The white and showed boxes represent functionality and mechanisms, respectively.} 
	\centering
	\label{fig:overview} 
\end{figure}

\name consists  of multiple components, shown in Figure~\ref{fig:overview}. The dynamic profiling component collects memory access information at the data object level, and decides the lifetime of data objects based on customized memory allocation and limited user annotation. The dynamic profiling only uses \textcolor{dong2}{one} training step to collect the information. After that, \name re-organizes memory allocation for short-lived data objects to facilitate data management and avoid page-level false sharing.

Driven by the profiling results, we treat short-lived and long-lived data objects separately. Short-lived data objects are allocated in a contiguous memory space in fast memory, and are not involved in data movement between fast and slow memories. This method avoids inefficient data movement due to short liveness. 

To handle long-lived data objects, \name uses an adaptive migration algorithm. The algorithm partitions the training process in a training step into migration intervals, based on the DNN model topology. In a migration interval, \name migrates data objects needed for the next interval, overlapping application execution with data migration. During data migration, \name must determine an appropriate migration interval, such that the data objects can be timely migrated from slow to fast memory before they are needed by application execution. We formulate the problem and determine the optimal migration interval. We also use a test-and-trial algorithm to determine if the migration cannot happen timely, whether continuing migration or not can lead to better performance. 

In general, \name uses the following domain knowledge to enable high performance of DNN training. 
\begin{itemize}
    \item Repetitiveness of DNN training for profiling and predicting memory access patterns;
    \item The liveness of data objects (tensors) within and across layers to decide data migration;
    \item The DNN model topology (i.e., layers) and \textcolor{dong2}{its depth} to decide the optimal migration interval and trigger data migration.
    %\textcolor{dong}{anything else?}
\end{itemize}

\subsection{Dynamic Profiling and Data Reorganization}
\label{sec:dyn_profiling_data_org}

%\textbf{Dynamic profiling.} 
\name integrates the profiling framework in Section~\ref{sec:profiling_framework} into the TensorFlow runtime system. We favor dynamic profiling instead of static one, although the static dataflow graph can be known before the training starts, because thread-level parallelism within an operation and across operations cannot be captured by static profiling. Such parallelism has significant impacts on data locality.  %Our profiling method uses \textcolor{red}{xxx\%} of total training time, which is a rather small overhead.

%textcolor{green}{We adopt the profiling flamework to get combined data access information in page-level and application-level for online profiling. The online profiling method is critical for the data migration on machine learning flamework since the data access are various between different parallelism strategy. Our online profiling method brings 4\% to 8\% runtime overhead. Leveraging the remarkable predictability of DNN model, Sentinel only needs to profile one training step(i.e, one mini-batch) to get data access information. Considering of DNN models usually needs tens of thousands of training steps to converge, the online profiling overhead is lightweight.}

%\textcolor{green}{All tensor allocation happens in slow memory during the profiling such that ensures there are enough space for tensor allocation. During the profiling, we maintain information of each data which used in one training step as data lifetime and size, and the access time in each operation.}\textcolor{blue}{Todo: why os level profiling with allocate each tensor pre-page.}

Based on the profiling results, \name uses a customized memory allocation for the remaining training steps. In particular, short-lived data objects that have the similar memory access pattern (including number of accesses and memory allocation and deallocation times) are allocated into the same page, in order to avoid page-level false sharing and reduce TLB misses. %\textcolor{dong}{Jie TODO: add more details here.}
\textcolor{dong2}{This is implemented by associating a bit string with each data object. The bit string indicates which layer this data object is accessed. Data objects that have the same bit string are grouped. Data objects falling into the same group are sorted in terms of number of memory accesses. 
The data objects in a group are allocated and packed into the same set of pages, following the increasing order.}

%%It would be better to elaborate a bit more on your algorithm of reorganizing the objects/page mapping.  Do you regroup short-lived tensors as well long-lived tensors? Do you maintain a queue internally to track the object allocation requests? Is it a ranking + greedy algorithm, does it consider padding?

%\textcolor{green}{\name reorganize the location for short-lived and long-lived data objects. Specifically, \name maintains a bit string to indicate whether data object be accessed in each DNN model layer. \name groups tensors with identical bit string together for allocation. \name allocates data objects in the same group in the same set of pages. }

Furthermore, \name preallocates a memory pool to meet the memory allocation requests for short-lived data objects. Since those data objects are frequently allocated and freed, using the memory pool can avoid repeatedly returning memory to the system, mitigating unnecessary overhead. 

%\textcolor{dong}{Memory pre-allocation.} \textcolor{green}{\name allocator pre-allocates heterogeneous memory for tensors for two reasons. Firstly, the default allocator in TensorFlow frequently allocates/deallocates large amount of tensors, which causes unnecessary runtime overhead. \name allocator holds on to released memory rather than releasing to OS directly for better performance. Secondly, the default allocator in TensorFlow allocates tensors without considering tensor access pattern. Allocator in \name allocates tensors with similar access pattern(i,e., tensors with similar access frequency, allocation and deallocation time) according to profiling result, which can reduce unnecessary page migrations and TLB miss.}

%\textcolor{dong}{Data reorganization.} \textcolor{green}{For tensors allocated during the profiling, \name reorganizes them according to profiling result. As Figure~ shows, most of tensors' lifetime is less than one training step. The runtime overhead for reorganizing tensors is ignorable.}

\subsection{Handling Short-Lived Data Objects}
%\textcolor{red}{Paragraph: define short-live tensors. }
%\textcolor{green}{\name considers tensor whose lifetime is shorter than one layer.}
%%We define short-lived data objects as those whose lifetime is within one layer. Short-lived data objects are allocated and freed within one layer. %Driven by our profiling results, we treat short-lived and long-lived data objects differently. This section discusses how to handle short-lived data objects, and the next section (Section~\ref{sec:adaptive_dm}) discusses how to handle long-lived data objects. 
During DNN training, a single short-lived data object is not accessed many times (e.g., less than 10 times in ResNet) in main memory, compared to many long-lived data objects. Hence, the data placement of a specific short-lived data object has an ignorable impact on the performance of DNN training. However, as our profiling results show that there are a large amount of short-lived data objects throughout the whole training process, and they share the same memory access characteristics (i.e., short-liveness, small size, and a small number of accesses in main memory). We must use a general policy to manage them.   

%\textcolor{red}{Paragraph: describe our algorithm that reserves a space in fast memory to handle small data objects}
We use the following algorithm to manage short-lived data objects. We allocate a continuous memory space in fast memory for short-lived data objects. Data objects in this space are never considered for migration. This space is reused for short-lived data objects, as they are allocated and freed throughout the training steps. \textcolor{dong2}{The space is allocated at the beginning of each migration interval to accommodate short-lived data objects in the interval. Doing this, \name guarantees that there is always memory space for short-lived data objects (i.e., no competition from long-lived data objects, because the placement of short-lived data objects is critical for performance.} 
%dynamically expanded, when there is no free space to accommodate the allocation requests;  
Within an migration interval, the space is  dynamically shrunk to free space for long-lived tensors, when a memory page in the space is freed. We collect short-lived data objects in this memory space, such that those short-lived data objects allocated and accessed at the similar time can be placed into the same page to avoid page-level false sharing.

%\textcolor{red}{Paragraph: the reason why we never move short-lived data objects out of fast memory. Mention that the existing algorithm can inappropriately handle short-lived data objects.}
\textcolor{dong}{The above method addresses the limitation of the existing methods that use a caching algorithm~\cite{Ramos:ics11, RAMinate:socc16, 5260554, Yan:ASPLOS19, heteros:isca17} or counting the number of memory accesses within a time window~\cite{Thermostat:asplos17, unimem:sc17}.} They  move short-lived data objects to slow memory, even though they are not accessed any more. This has two problems: (1) Unnecessary data movement causes performance loss and wastes memory bandwidth; (2) Short-lived data objects unnecessarily stay longer in fast memory, wasting valuable fast memory space. This is because making the decision on the movement of short-lived data objects takes some time, due to the necessity of collecting memory access information to run the caching algorithm. In addition, counting the number of memory accesses for individual data objects can be inaccurate, because they can share memory pages and the number of memory accesses to each data objects is small. Using our algorithm based on the DNN domain knowledge, we do not have the above limitation.

%\textcolor{red}{Paragraph: what if there is no enough space to hold short-lived data objects? This case will never happen.}
In our design, fast memory is always large enough to host short-lived data objects. If not, short-lived data objects will be frequently moved between fast and slow memories. This data movement is highly inefficient in terms of both performance and energy efficiency, especially for data objects with a short lifetime. Hence, we assume that the fast memory size is at least larger than the peak memory consumption of those short-lived data objects. We have a discussion on the fast memory size in Section~\ref{sec:discussion}.  

Since short-lived data objects are frequently allocated and freed and we reuse the same memory space to host them, the size of memory space for short-lived data objects is small, and typically bounded by a few GB.

\subsection{Adaptive Data Migration}
\label{sec:adaptive_dm}

\begin{figure}
\centering
\includegraphics[width=0.48\textwidth]{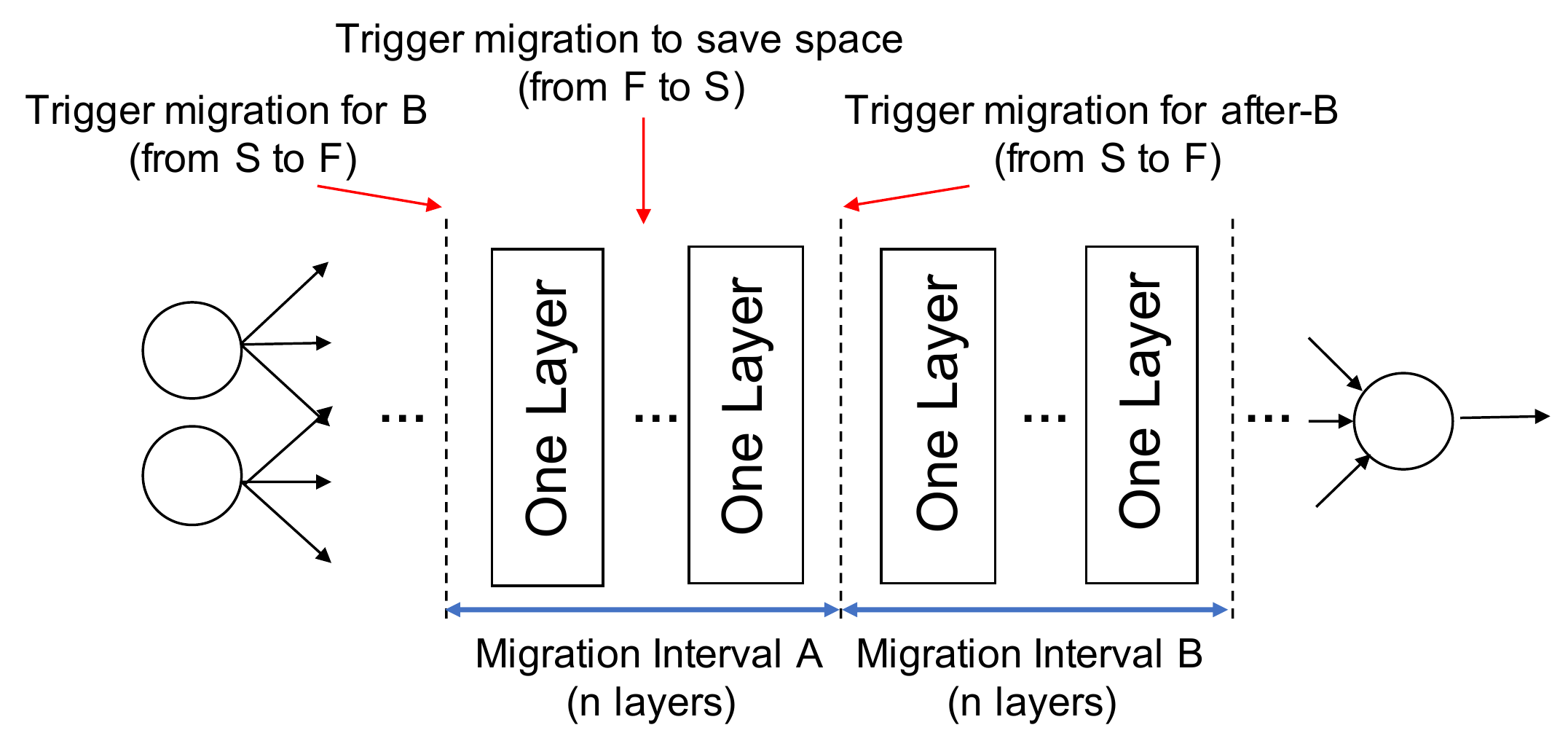}
	\caption{Data migration based on the migration interval. ``S'' and ``F'' stand for slow and fast memories respectively.} 
	\centering
	\label{fig:migration_interval} 
\end{figure}

%\textcolor{dong}{paragraph: basic idea about the data migration and migration interval. Add a figure here?}
We migrate data for those long-lived data objects. %and short-lived tensors (if the space of fast memory is not enough to accommodate the upcoming data migration). 
The data migration is controlled by the \textit{migration interval}. The migration interval determines how frequently we migrate data between fast and slow memories. A training step is partitioned into many equal-sized
migration intervals. Figure~\ref{fig:migration_interval} generally depicts data migration.

Data migration from slow to fast memory is triggered at the beginning of each interval, aiming to prefetching data objects needed by the next interval into fast memory before the next interval starts. The data migration happens in the middle of each interval, in order to overlap data migration with DNN training as much as possible, such that the overhead of data migration is removed from the critical path. 

Data migration from fast to slow memory is triggered and happens in the middle of the interval, when the long-lived data object is not accessed in the interval. Such data migration is used to save the space of fast memory as much as possible, in order to accommodate upcoming data migration. %and save space for short-lived data objects.
%\textcolor{green}{For each migration interval, \name migrates data into fast memory which will be used in next migration interval execution. Using migration interval ensures \name migrate data for future usage.}

\begin{comment}
\begin{figure*}[htb!]
	\centering
	\includegraphics[width=1.0\linewidth, height=0.2\textheight]{./figures/migration_interval.pdf}
	\caption{Data migration based on the migration interval. ``S'' and ``F'' stand for slow and fast memories respectively.} 
	\centering
	\label{fig:migration_interval} 
\end{figure*}
\end{comment}

%\textcolor{red}{paragraph: the migration interval is defined in terms of layers. why?}
We define the migration interval in terms of layers in DNN, not in terms of execution time, because of the following three reasons. First, the layer-based migration interval naturally guarantees the completion of operations at the end of the interval, because no operation runs across layers. The time-based migration interval cannot guarantee that, which brings inevitable synchronization between application execution and data migration, causing performance loss. Using the DNN domain knowledge (i.e., the layers), we avoid the above problem. Second, each layer is associated with a computation phase that shows a memory access pattern. The layer-based migration interval allows us to easily leverage the memory access patterns collected at the profiling phase to guide data migration. Third, the time-based migration imposes challenges on deciding which operations are in which migration interval, because of operation-level parallelism.

\begin{figure}
	\centering
	\includegraphics[width=0.48\textwidth]{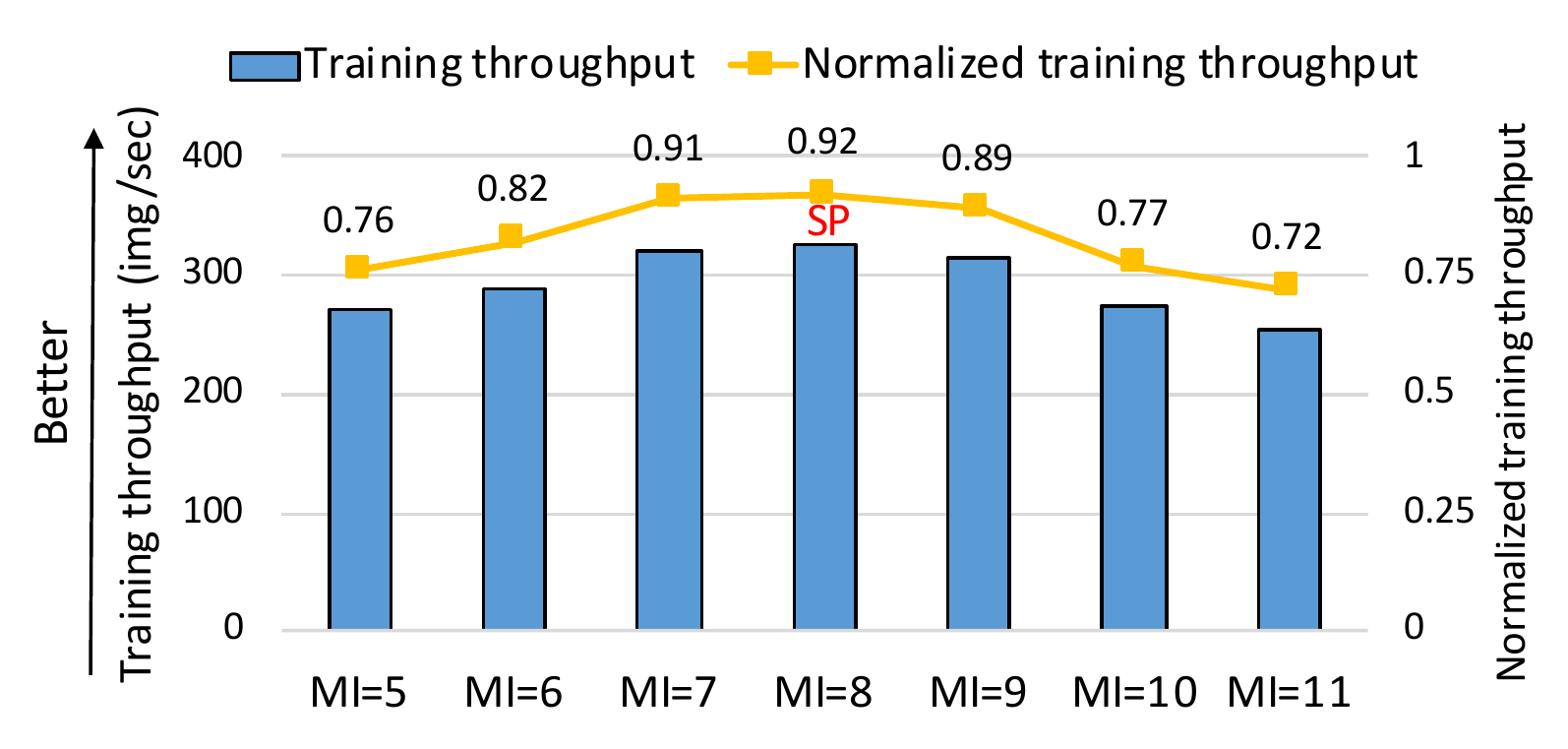}
\caption{Performance (training throughput) variance as we change the migration interval (MI). ``SP'' stands for sweet spot (the optimal migration interval).}%\textcolor{green}{Todo}}
	\label{fig:optimal_migration_interval}
\end{figure}

\textbf{Determining an appropriate migration interval} is challenging. If the migration interval is either too large or too small, we cannot achieve the best performance. %\textcolor{red}{paragraph: show the sweet spot of migration interval with figures} 
%To determine the existence of an optimal migration interval, we use different migration intervals and then measure the performance. 
\textcolor{dong}{
Figure~\ref{fig:optimal_migration_interval} shows the performance when we use different migration intervals for training ResNet\_v1-32 \textcolor{dong}{with 1GB fast memory}. The figure reveals that the performance is very sensitive to the migration interval. There is 21\% performance variance when we change it from 5 to 11. When the migration interval is 8, we achieve the best performance. Hence, determining an appropriate migration interval is critical for performance.}

We analyze the trade-off between large and small migration intervals as follows. 
If the migration interval is large, then the data to migrate for this interval is large. The migration interval cannot be too large. Otherwise the data to migrate can be larger than the available space in fast memory. This constraint on the migration interval is the \textit{space constraint}, formulated in Equation~\ref{eq:space_constraint}.

If the migration interval is small, then the available execution time to overlap data migration with application execution is short. The migration interval cannot be too short. Otherwise the data to migrate cannot be timely migrated from slow to fast memory before the next migration internal starts. This constraint on the migration interval is the \textit{time constraint}, formulated in Equation~\ref{eq:time_constraint}.

In Equations~\ref{eq:space_constraint} and~\ref{eq:time_constraint} , $RS$ is the fast memory space for short-lived data objects, $S$ is the fast memory size, and $MI$ stands for the migration interval. $RS$ is a function of the migration interval (different migration intervals have different $RS$). In Equation~\ref{eq:space_constraint}, $Data$ is the size of data for migration in a migration interval; In Equation~\ref{eq:time_constraint}, $BW$ is the migration bandwidth from slow to fast memory, and $T$ is the DNN training time in a migration interval. \textcolor{dong}{$Data$ and $T$ are functions}
% $T$ is a function 
of the migration interval (different migration intervals have different \textcolor{dong}{$Data$ and }$T$).  

%In both cases, we increase the risk of unfinished data migration before the next migration internal happens. 
%The two constraints establish the upper and lower bounds on the migration interval. There is an optimal migration interval, with which we can maximize the overlap between data migration and application execution and hence minimizes data migration cost.
%\vspace{-10pt}
\begin{equation}
%\vspace{-10pt}
\label{eq:space_constraint}
   \text{Space constraint:} \quad Data(MI) < S - RS(MI)  
   %\vspace{-10pt}
\end{equation}

\begin{equation}
\label{eq:time_constraint}
   \text{Time constraint:} \quad T(MI) > (S - RS(MI))/BW 
%\vspace{-5pt}
\end{equation}

$RS$ is relatively stable, according to our profiling results. There is a small variance as we change $MI$. Hence $S - RS(MI)$ is near constant. $Data(MI)$ and $T(MI)$ are monotonically increasing functions of $MI$ (i.e., a larger $MI$ indicates larger $Data$ and $T$, and vice versa).
Hence, the two equations establish the upper and lower bounds on the migration interval. 

The two equations, although revealing the inherent trade-off between small and large migration intervals, cannot reveal the optimal one, because they do not capture the data movement from fast memory to slow memory. Such data movement increases the available fast memory space. Because of such data movement, those migration intervals that meet the two constraints can perform differently. 

We use the following method to determine the optimal migration interval at runtime. After collecting the profiling results, we use Equations~\ref{eq:space_constraint} and~\ref{eq:time_constraint} to prune the search space of the migration interval and choose those that meet the constraints. Then we use a few more training steps, each of which employs a migration interval. We measure their performance, and choose the optimal migration interval that leads to the best performance.

%\textcolor{red}{paragraph: discuss the three possible cases at the end of the migration interval.}
\textbf{We encounter three possible data migration cases} at the end of a migration interval. We discuss them as follows. Assume that we have two intervals, $A$ and $B$, and $B$ is right after $A$. \name migrates data at the beginning of $A$ for $B$. At the end of $A$, we have three cases. %The three cases can happen, even if we use the optimal migration interval.
\begin{itemize}
    \item Case 1: All data migration has been finished;
    \item Case 2: Data migration cannot be finished, because fast memory cannot offer enough free space;
    \item Case 3: Data migration cannot be finished, because there is no enough time for migration (there is still space in fast memory).
\end{itemize}

\begin{figure}[htb!]
	\centering
	\includegraphics[width=0.48\textwidth]{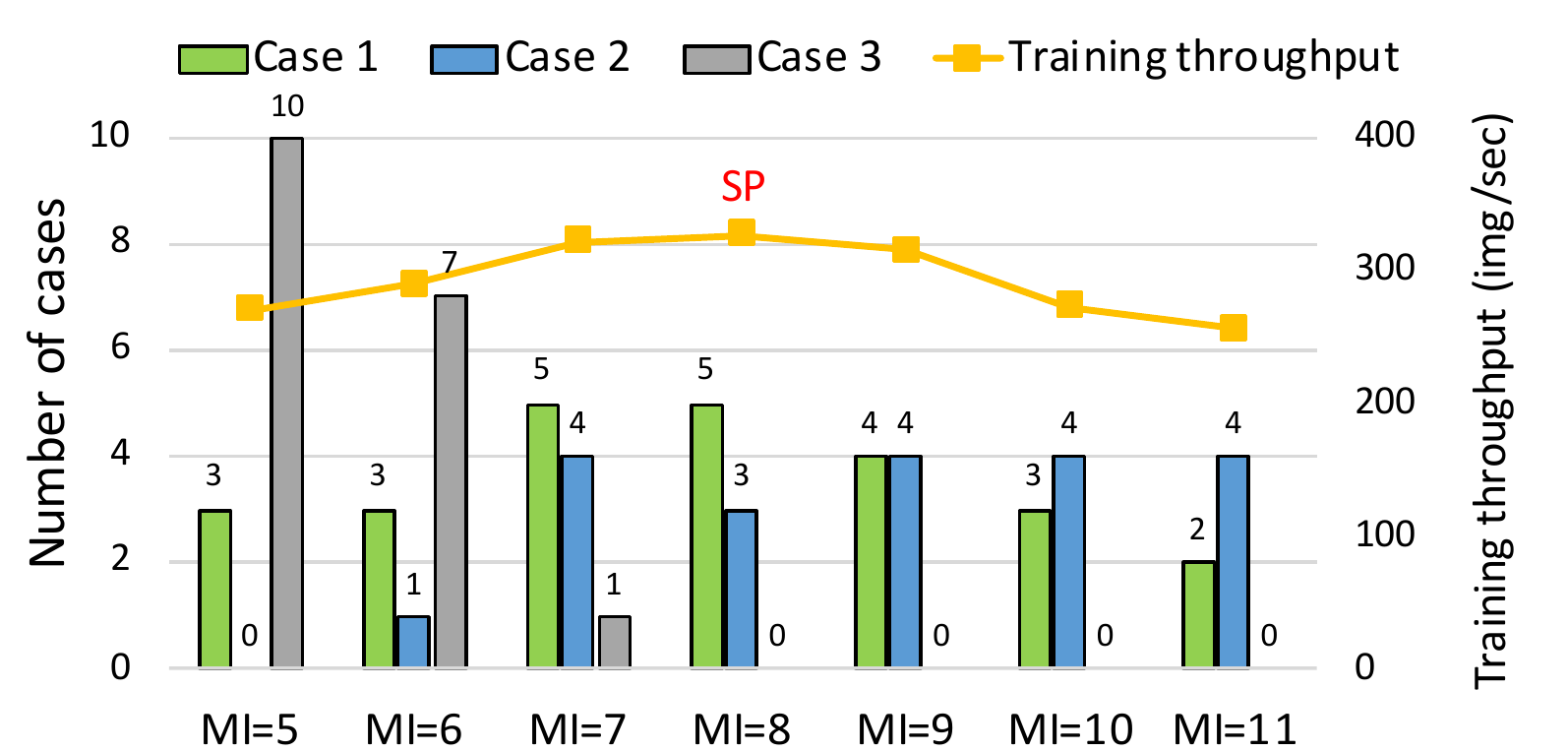}
    \caption{The occurrences of the three data migration cases in a training step of ResNet-32. ``MI'' stands for ``migration interval''. "SP" stands for sweet spot (the optimal migration interval). %\textcolor{green}{TODO: remove percentage, MI=8}
    }
    \label{fig:migration_interval_three_cases}
\end{figure}

In Case 1, once $B$ starts, all of the migrated data object are in fast memory, which is the ideal case. For Cases 2 and 3, we must avoid them. %\textcolor{green}{show the frequency of three possible cases at the end of the migration interval with different migration interval?}
The migration interval has impact on how often the three cases happen. \textcolor{dong}{Given a specific fast memory size, a} small interval can create more Case 3, and a large interval can create more Case 2. Figure~\ref{fig:migration_interval_three_cases} shows how many times each case happens, when we use different migration intervals for training ResNet\_v1-32 \textcolor{dong}{with 1GB fast memory}. \textcolor{dong}{When the migration interval decreases from 11 to 5, Case 3 increases from 0 to 13; When the migration interval increases from 5 to 11, Case 2 increases from 0 to 4. This result is consistent with our analysis.}

%The figure shows that even though we choose the optimal migration interval, \textcolor{red}{Cases 2 and 3 still happen.} 

%\begin{comment}

To avoid Case 2, long-lived tensors are immediately moved out of fast memory in the middle of $A$, once the remaining operations in $A$ do not need them. This saves space of of fast memory. However, avoiding Case 3 is difficult, because it is created by the limited memory bandwidth and/or latency. 
In Case 3, we can either continue migrating data and let $B$ wait for the completion of data migration, or leave data in slow memory. The continuation of data migration exposes data migration into the critical path, but the execution of $B$ use data in fast memory; On the contrary, leaving data in slow memory uses the data in slow memory but avoids data migration overhead. This is a classic trade-off between data locality and data movement. To determine which method leads to the best performance, we use a test-and-trial algorithm.

In particular, whenever Case 3 happens at the end of an interval, we use one training step to try the continuation of data migration, and use another training step to try no-data-migration. We measure the performance of the two methods and use the best method in the remaining training steps. Note that in order to compare the performance of the two training steps, we must ensure that data placement in the two training steps is the same when Case 3 happens. The same data placement can be easily guaranteed, given the repetitive execution pattern in DNN training. 

The above test-and-trial algorithm does not cause large overhead, because Case 3 does not happen often and hence does not need a large amount of training steps for test and trial. \textcolor{dong2}{The number of training steps used in test and trial is usually less than 10 (see Table~\ref{tab:models}). }
%two times of the number of migration intervals

%In practice, we find that \textcolor{red}{xxx} training steps are enough for DNN models in our evaluation. 

\subsection{Discussions}
\label{sec:discussion}

\textbf{The lower bound of fast memory size.} Although fast memory can be smaller with \name, there is a lower bound of fast memory size to avoid big performance loss. This lower bound is the peak memory consumption of short-lived data objects in any migration interval plus the largest long-lived data object. Smaller than this lower bound, the runtime system has to either frequently migrate short-lived data objects or has no space to accommodate long-lived data objects, which usually causes performance loss larger than 10\%.

\textbf{Handling dynamic graphs.}
Some machine learning frameworks, such as PyTorch and TensorFlow 2.0, support dynamic graphs. Depending on the size of input within a  mini-batch, these frameworks generate a different graph with the right shape to accommodate the mini-batch. With dynamic graphs, mini-batches are not identical. Hence, there could be multiple dataflow graphs. 

To handle dynamic graphs, the existing solution pads zero at the end of input~\cite{google_tensorflow_bucketing}, such that mini-batches have the same structure. This transforms a dynamic graph into a static one, but at the cost of larger memory footprint and unnecessary computation. We use a solution similar to the one in~\cite{DBLP:conf/asplos/SivathanuCSZ19} that uses bucketed profiling. In particular, \name bucketizes the input sizes into a small of buckets (at most 10 in \name), and each bucket has a similar graph. \name profiles each bucket to collect memory access information and decide data migration.

\textbf{Handling control dependencies.}
A static graph can have control flow. Depending on the value of input in a mini-batch, the graph can have different dataflow, causing different memory access patterns. \name handles this case by tracking dataflow. Whenever a new dataflow is encountered, \name triggers profiling and makes the decision of data migration again.

%% file: text/impl.tex
\section{Implementation}
\label{sec:impl}

We implement \name in Linux v4.9 and TensorFlow v1.14. We change the Linux kernel for memory profiling; We change the TensorFlow runtime system for page migration. The statistics of kernel modification given by \texttt{git diff} is \textcolor{dong}{17 files changed, 587 insertions(+), 18 deletions(-)}; the statistic of TensorFlow modification given by \texttt{git diff} is \textcolor{dong}{33 files changed, 2425 insertions(+), 37 deletions(-)}.

\name introduces three APIs to trigger/stop memory profiling and identify layers, which are \texttt{start\_profile()}, \texttt{end\_profile()}, and \texttt{add\_layer()}. \texttt{start\_profile()} triggers a system call to enable tracking main memory accesses, and enables tracking of memory allocation/deallocation to record lifetime information for data objects. \texttt{add\_layer()}, placed at the end of each layer, informs the runtime system of where is each layer to determine migration interval. Adding \texttt{start\_profile()}, \texttt{end\_profile()} includes only two lines of changes to the DNN model. Adding \texttt{add\_layer()} includes \~10-100 lines, depending on how many layers there are in the DNN model. Adding those APIs do not impact execution correctness of DNN training.

%\textcolor{green}{ \name expects three user annotations to trigger profile and page migration for DNN models. These annotation are implemented as TenorFlow operations and provided Python interface. And all of the annotations do not effect the computation of DNN models \texttt{start\_sentinel\_profile()} and \texttt{end\_sentinel\_profile()} specify the profiling step. Those APIs trigger system call to track main memory page access. Meanwhile those APIs enables tracking TensorFlow default allocater/deallocator to record tensor lifetime information.  Annotation \texttt{sentinel()} marks the minimum migration granularity of \name. \name expects user to place this annotation at the end of each DNN layer. Note that \name guarantees execution correctness regardless of where annotations are placed. }

\begin{figure}
\centering
%\vspace{-20pt}
\includegraphics[width=0.48\textwidth]{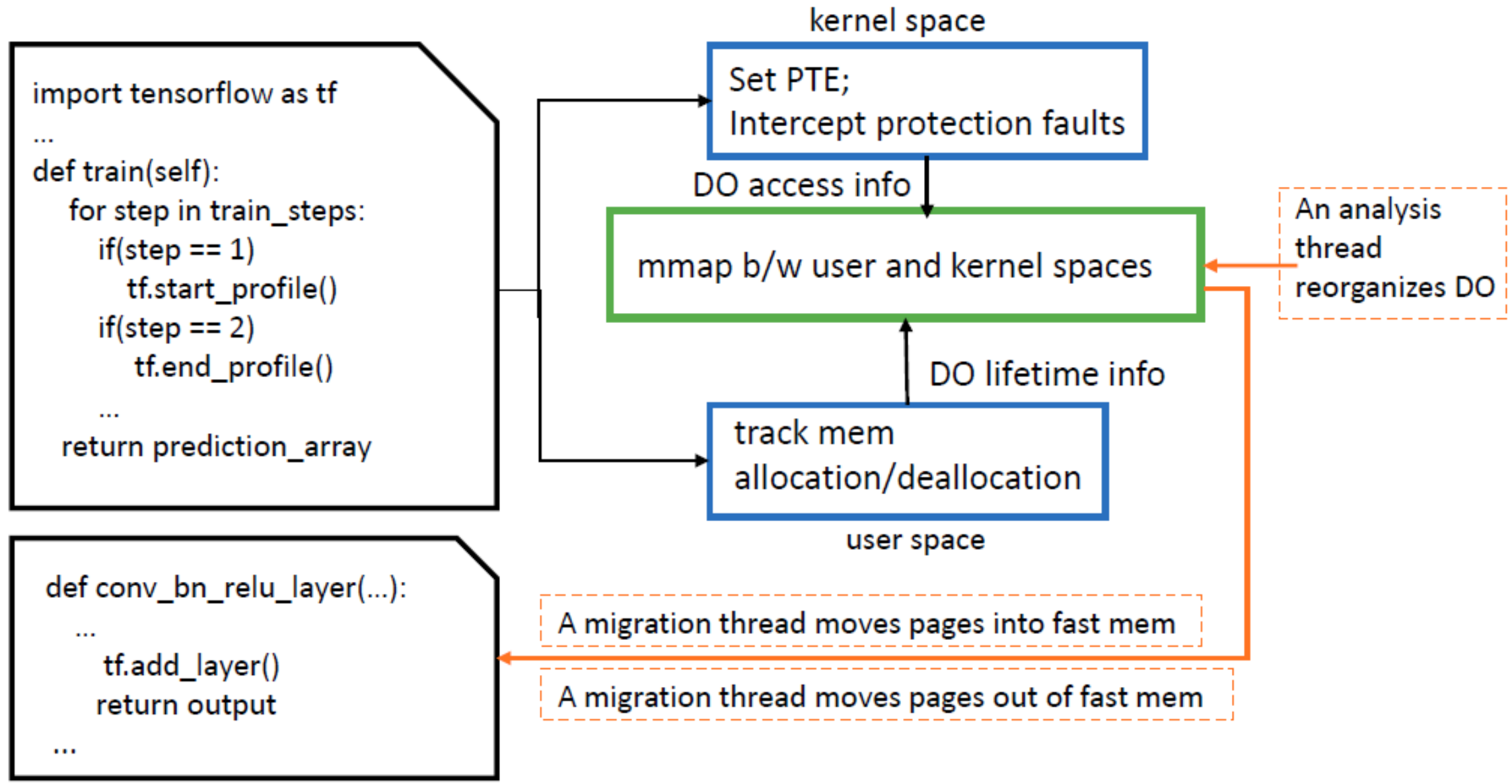}
%\vspace{-25pt}
\caption{Overview of \name implementation. ``DO'' in the figure stands for ``data objects''.}
%\vspace{-10pt}
\label{fig:impl}
\end{figure}

%\textcolor{red}{Add a figure to explain the impl. The figure includes the user annotation, the helper threads for data migration, and the runtime thread to analyze profiling results and re-organize data.}

%\textcolor{green}{ \name implemenation includes kernel modification and TensorFlow Runtime modification.
%}

Figure~\ref{fig:impl} shows some implementation details. After collecting memory access information from OS and lifetime information from the TensorFlow runtime, \name issues three helper threads: one for information analysis to determine migration interval and making migration decision, one for data migration from fast to slow memory, and one for the migration in the opposite way. The two migration threads work in parallel to accelerate migration. 
\textcolor{dong}{\name uses the Linux system call \texttt{move\_pages()} to migrate pages.}

%\textcolor{green}{Figure~\ref{fig:impl} shows \name implementation details. \name gets data object access information from OS and data object lifetime information from TensorFlow Runtime. \name manages kernel mmap space for user and kernel space communication. \name allocates each data object in at least on page to obtain data object access information during the profiling. After the profiling, \name leverages a analysis thread to handle false sharing data objects and reorganize data objects allocated during the profiling step. Meanwhile \name uses two threads for page migration between fast memory and slow memory.}

%% file: text/evaluation.tex
\section{Experimental Results}
\subsection{Methodology}

We study HM in a machine with two memory nodes. We use one as fast local memory and one as slow remote memory. Table~\ref{tab:hardware} summarizes the hardware we use. %for evaluation. %We inject traffic by using memhog inject traffic to change the bandwidth of slow memory.

%\textcolor{green}{We implement \name in Linux v4.9 and TensorFlow v1.14. The kernel change is for profiling and the Tesorslow change is for runtime page migration. The statistics of kernel modification given by git diff is X; the statistic of TensorFlow source code modification given by git diff is X.}

%\textcolor{green}{We evaluate 5 DNN models to verify the effectiveness of \name. Table~\ref{tab:models} shows the detail information. We use the implemetation of ResNet\_v2-152, LSTM, and Wide\_Deep model from TensorFlow software package~\cite{tf_models}, ResNet-32 from ~\cite{dcgan} and DCGAN from~\cite{resnet_32}.  The default intra-op and inter-op parallelisms are set as the number of one socket physical cores of the hardware platform (24 on the experimental platform).}

We evaluate five DNN models.
%to verify the effectiveness of \name. 
Table~\ref{tab:models} shows model details. For ResNet\_v2-152, LSTM, and %Wide\_Deep 
\textcolor{dong}{MobileNet} models in our evaluation, we use the implementations from TensorFlow~\cite{tf_models}; For ResNet\_v1-32 and DCGAN, we use~\cite{resnet_32} and~\cite{dcgan} respectively. %For DCGAN, we use~\cite{dcgan}.  
To use TensorFlow, the intra-op parallelism (i.e., the number of threads to run an operation) and inter-op parallelism (i.e., the maximum number of operations to co-run) are set as 24, which is the number of physical cores in a socket in our platform. 

We compare \name with a state-of-the-art page migration system from Yan et al.~\cite{Yan:ASPLOS19}. They introduce a page migration algorithm based on an existing page replacement mechanism in the Linux kernel (i.e., the FIFO-based active list~\cite{Yan:ASPLOS19}). In~\cite{Yan:ASPLOS19}, they improve the performance of the page migration mechanism by using four threads for parallel page copying and eight threads for concurrent page migration, and they optimize page locations every five seconds. We use the same configuration in our evaluation. \name does not use the page migration mechanism in~\cite{Yan:ASPLOS19}. \textcolor{dong}{Unless otherwise indicated, the size of fast memory in our evaluation is equal to 20\% of peak memory consumption in DNN models.}

%\textcolor{green}{We compare \name with the state-of-the-art multi-level page migration system from Yan et al.~\cite{Yan:ASPLOS19}. Yan proposes a page migration algorithm builds on existing Linux kernel page replacement algorithm. We mention Yan's page migration algorithm as \baseline. \name migrates pages according to page's state, page table entry’s access bit and page metadata access bit maintained by kernel. We set \baseline using 4-thread parallel and 8 page concurrent migration, and set \baseline optimizing page locations every 5 seconds. The configuration is based on \baseline optimal solution according to ~\cite{Yan:ASPLOS19}. }

\begin{table}[t]
\centering
\scriptsize
%\small
%\vspace{-3pt}
%\caption{\textcolor{dong}{Hardware used in evaluation.}}
\caption{Hardware overview of experimental system.}
%\vspace{-11pt}
\begin{tabular}{ll}
\hline
CPU& 2-socket Intel(R) Xeon(R) CPU E5-2670 v3 \\ %\phantom{LA}  @2.30GHz        
DRAM& DDR4 - 2133MHz   \\
Fast memory & BW: 34 GB/s  \phantom{L}  Latency: 87 ns \\ % \phantom{LA}    
Slow memory& BW: 19 GB/s  \phantom{L}  Latency: 182.7 ns\\
Cross-socket BW & 19 GB/s\\ 
\hline
\end{tabular}
%\vspace{-25pt}
\label{tab:hardware}
\end{table}

\begin{comment}
\begin{table*}[h]
\centering
%\small
\caption{DNN models for evaluation. ``p and t'' stands for profiling and test-and-trial} 
\begin{tabular}{|l|l|l|l|l|}

\hline
               & data set      & batch size & \# of training steps for ``p and t'' & total \# training steps \\ \hline
ResNet\_v1-32   & CIFAR-10      & 128  & 8 &  80K\\ \hline
ResNet\_v2-152 & CIFAR-10      & 32  &   &  80K\\ \hline
LSTM           & PTB           & 20   & 2 & 5500    \\ \hline
DCGAN          & MNIST         & 64  &   &   \\ \hline
MobileNet     & CIFAR-10 & 64 &                              &                      \\ \hline
\end{tabular}
\label{tab:models}
\end{table*}
\end{comment}

\begin{table}[h]
\centering
\small
\caption{DNN for evaluation. ``p, m \& t'' stands for profiling, determining optimal migration interval, and test-and-trial.} 
\begin{tabular}{|c|c|c|c|}

\hline
               & data set      & \begin{tabular}[c]{@{}c@{}}batch \\ size\end{tabular} & \begin{tabular}[c]{@{}c@{}}\# of training \\steps for ``p,m \& t''\end{tabular}  \\ \hline
ResNet\_v1-32   & CIFAR-10      & 128  & 8 \\ \hline
ResNet\_v2-152 & CIFAR-10      & 32  &  5 \\ \hline
LSTM           & PTB           & 20   & 2   \\ \hline
DCGAN          & MNIST         & 64  & 4  \\ \hline
MobileNet     & CIFAR-10 & 64 &      3    \\ \hline
\end{tabular}
\label{tab:models}
\end{table}

%\textcolor{red}{Add a table to show (1) benchmark names; (2) training data name; (2) batch size; (3) the total number of training steps for profiling and test-and-trial, and (4) the total number of training steps.}

\subsection{Results}
\begin{figure}
    \centering
    \includegraphics[width=0.48\textwidth]{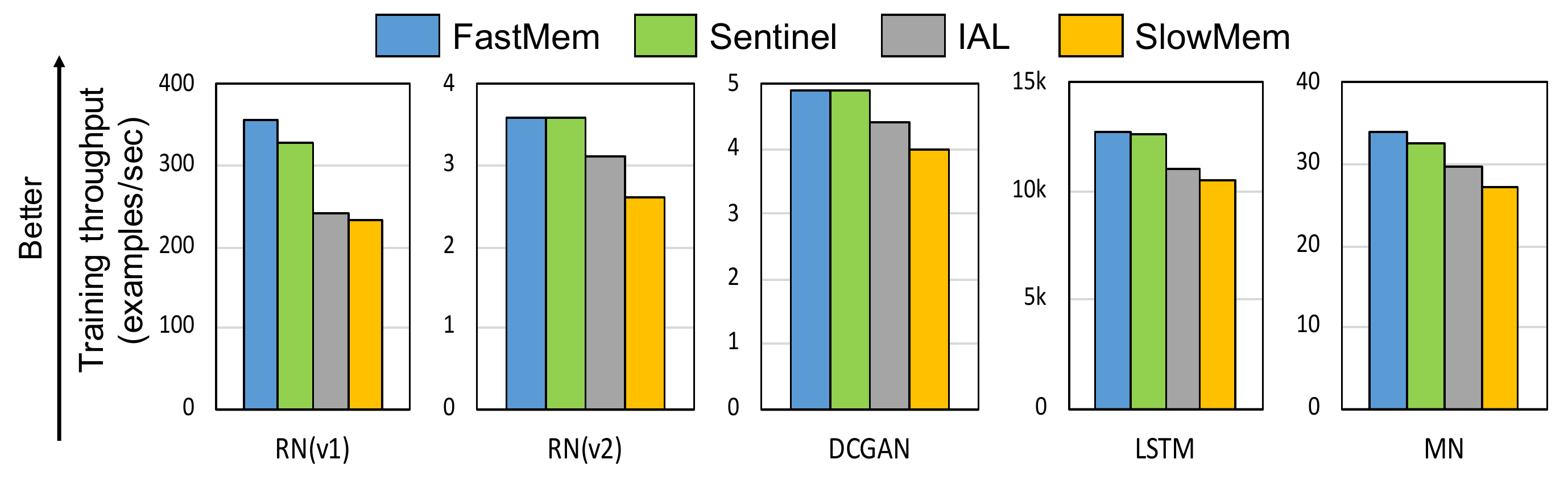}
    
\caption{Performance with \name, IAL and two  configurations. \textcolor{dong}{``RN(v1)'', ``RN(v2)'', and ``MN'' stand for ``ResNet\_v1-32'', ``ResNet\_v2-152'', and ``MobileNet'', respectively.}}
\label{fig:general_perf} 
\end{figure}

\textbf{Overall performance.}
Figure~\ref{fig:general_perf} shows performance of \name and compare it with the improved active list (IAL) for HM in~\cite{Yan:ASPLOS19} (a state of the art). The figure shows that performance difference between \name and the fast memory-only system is very small (no difference in \textcolor{dong}{two} models and at most \textcolor{dong}{8\%} difference in \textcolor{dong}{ResNet\_v1-32}), while IAL has \textcolor{dong}{17}\% performance difference on average (up to \textcolor{dong}{32}\%). \name is significantly better than IAL by \textcolor{dong}{18\%} on average (up to \textcolor{dong}{37\%}).

Table~\ref{tab:migration} shows the number of migrations in \name and IAL. Compared with IAL, \name has more migrations (\textcolor{dong}{88\% more on average}). Frequent migrations allow \name to make best use of fast memory for performance; Also, those migrations are successfully overlapped with DNN training to avoid performance loss.  

%Such reduction comes from reducing unnecessary data movement for short-lived data objects. For some long-lived data objects, test-and-trial also helps to reduce their data movement if the data movement is not helpful for performance. 

\pgfplotstableread{
Criterion   AA   BB   CC   DD   EE
C1   3.58  3.58 3.58 3.58  3.58
C2   3.58  3.58 3.58 3.58 3.58
}\migrationfdatatable

\begin{table}[]
\small
\caption{Number of page migrations in one epoch. ``RN(v1)'', ``RN(v2)'', and ``MN'' stand for ``ResNet\_v1-32'', ``ResNet\_v2-152'', and ``MobileNet'' respectively.}
\begin{tabular}{|p{1cm}|p{0.9cm}|p{1cm}|p{1cm}|p{0.9cm}|p{0.9cm}|}
\hline
   & RN(v1) & RN(v2) & DCGAN & LSTM & MN \\ \hline
IAL  & 807308	& 3432254 &	211684	&194933  & 144882 \\ \hline
Sentinel & 2097152	& 4898697	& 444846 &	353500  & 249290 \\ \hline
\end{tabular}
\label{tab:migration}
\end{table}

%\textcolor{red}{Add a table to show memory consumption}
Table~\ref{tab:mem_sentinel} shows \textcolor{dong}{peak} memory consumption before and after using \name. Although our profiling method %does not put more than one data objects into the same page for high profiling accuracy,
increases memory consumption, it does not increase much (by \textcolor{dong}{2.1\%} at most). This is because data objects larger than 4KB dominate total memory consumption. During the profiling, we do not significantly increase their memory consumption. 
%Hence the total memory consumption is no increased very much. 

\begin{table}[]
\caption{\textcolor{dong}{Peak} memory consumption with and without \name.}
\centering
\begin{tabular}{|l|l|l|}
\hline
                & w/o Sentinel & w/ Sentinel \\ \hline
ResNet\_v1\_32  & 6144 MB         & 6176 MB    \\ \hline
ResNet\_v2\_152 & 25600 MB         &  25856 MB  \\ \hline
LSTM            & 2048 MB   & 2080 MB            \\ \hline
DCGAN           & 3072 MB   &  3136 MB           \\ \hline
MobileNet       & 4096 MB       &  4228 MB          \\ \hline
\end{tabular}
\label{tab:mem_sentinel}
\end{table}

\begin{figure}[htb!]
	\centering
	\includegraphics[width=0.48\textwidth]{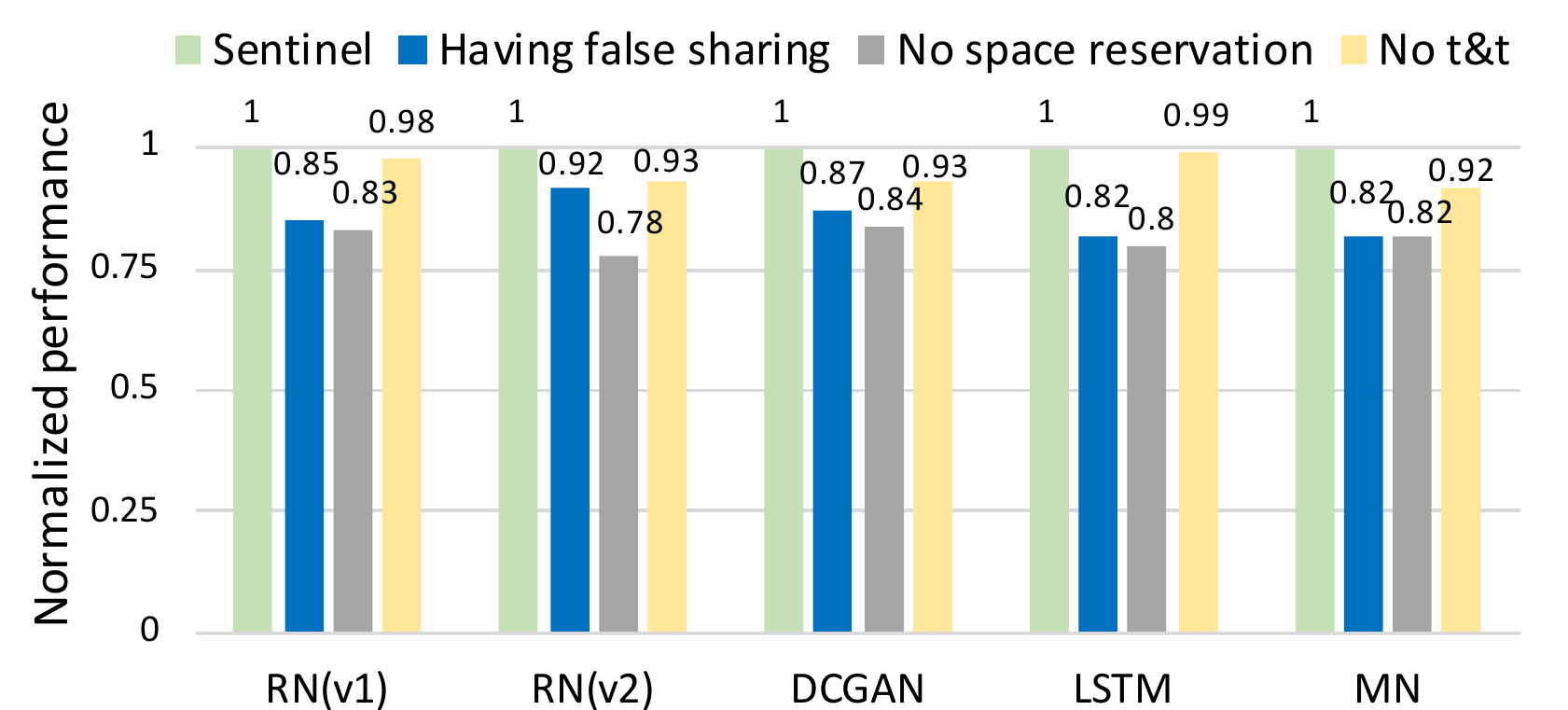}
\caption{\textcolor{dong}{Performance with different strategies for data management in \name. ``RN'' and ``MN'' stand for `ResNet' and  ``MobileNet'' respectively. Performance is normalized by the performance of the full-featured \name.}}
\label{fig:fast_memory}
\end{figure}

\textbf{Performance breakdown.}
%%\textcolor{red}{Figure: Multiple strategies, %showing memory pre-allocation, Sentinel without handling false sharing, no preservation space for short-lived variables, test and trail}
%\textcolor{green}{four bars groups together, 5 groups in one figure.} %\textcolor{red}{Figure: show how often the three cases happen at the end of each migration interval.}
We apply different strategies for data management, in order to study the impact of various techniques. Figure~\ref{fig:fast_memory} shows the results. In the figure, we show four strategies: \name without handling page-level false sharing (labeled ``Having false sharing''), \name without reserving fast memory space for short-lived data objects (labeled ``No space reservation''), \name without test-and-trial (labeled ``No t\&t''), and \name with all techniques.

The figure reveals that among the three (handling page-level false sharing, reserving fast memory, and test-and-trial), 
\textcolor{dong}{reserving fast memory space for short-lived data objects} is the most effective one. We easily have 17\% - 23\% performance loss without it (compared with the full-featured \name). Furthermore, because of the pervasiveness of page-level false sharing, handling false sharing improves performance by 8\% - 18\%.

\begin{figure}[htb!]
	\centering
	\includegraphics[width=0.48\textwidth]{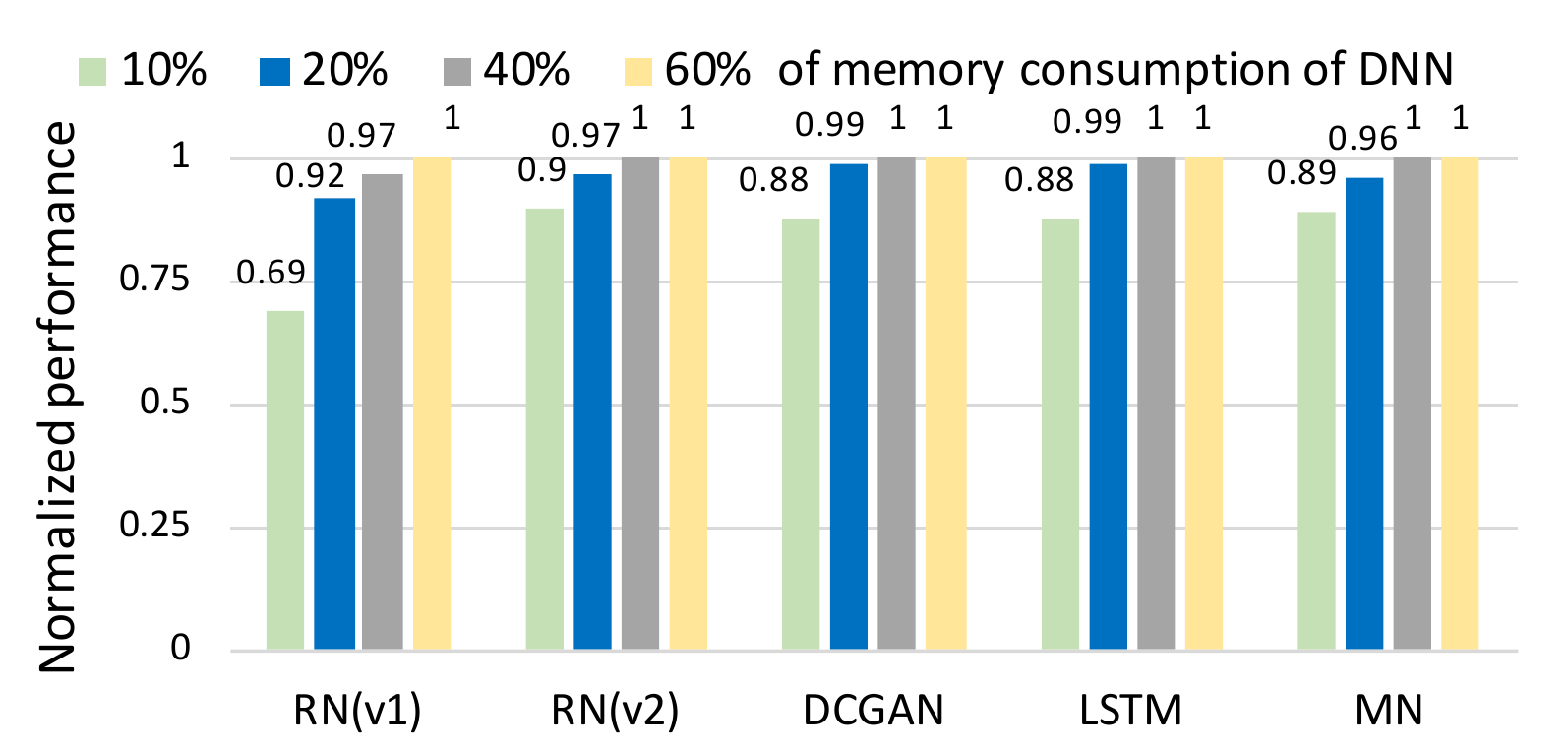}
\caption{Performance with \name under various sizes of fast memory. The fast memory size is shown as the percentage of \textcolor{dong}{peak} memory consumption of DNN models. Performance is normalized by that of the fast memory-only.}
\label{fig:fast_memory_sen}
\end{figure}

\textbf{Sensitivity study.}
%\textcolor{red}{Figure: sensitivity study: show the effects of fast memory capacity.}
%\textcolor{green}{5 bars groups together, 5 groups in one figure.} 
We change fast memory size and measure performance. Figure~\ref{fig:fast_memory_sen} shows the results. In general, larger fast memory gives better performance. When the fast memory size is \textcolor{dong}{60}\% of \textcolor{dong}{peak memory consumption}, all of DNN models with \name on HM do not have any performance difference from the fast memory-only system. Also, with \name, performance is not sensitive to fast memory size: There is only \textcolor{dong}{at most 8}\% performance variance when the fast memory size is changed from \textcolor{dong}{20\%} to \textcolor{dong}{40\% of peak memory consumption of DNN}. This result is a demonstration of how \name effectively uses data movement to make best use of fast memory. 

\textbf{Saving fast memory size.} 
Figure~\ref{fig:general_perf} shows that using 20\% of \textcolor{dong}{peak memory consumption}
%peak memory  consumption 
of DNN models as fast memory size, \name on HM has almost the same performance (8\% difference at most) as the fast memory-only. This brings 80\% saving in fast memory size. 
Figure~\ref{fig:fast_memory_sen} shows that using 60\% of \textcolor{dong}{peak memory consumption} of DNN models as fast memory size, there is no performance loss, which comes with 40\% saving in fast memory size.

\begin{comment}
\pgfplotstableread{	
Criterion AA BB
1 6 1.2
2 9 2
3  26 6.25
4 35 7.5
}\linetwofdatatable

\begin{figure}
\centering
\resizebox{0.4\textwidth}{0.2\textwidth}{
\begin{tikzpicture}
	\begin{axis}[
		height=6.5cm,
		width=8.5cm,
		grid=major,
		legend style={at={(0.5,1.25)},anchor=north},
	%	xlabel = Fast Memory Size (GB),
    ylabel = {Peak memory consumption (GB)},
   % ylabel style={text height=0.02\textwidth,xshift=-0.005\textwidth,inner ysep=0pt},
    ymin=0,ymax=40,
    legend columns=1,
    legend style={/tikz/column 2/.style={column sep=5pt},
    font=\small
    },xtick=data,xticklabels={ResNet(32),ResNet(56),ResNet(110),ResNet(152)}
    %yticks={0,10,20,30,40}
	]
\addplot table [y=AA] {\linetwofdatatable};
	\addlegendentry{Peak memory consumption in ResNet}
\addplot table [y=BB] {\linetwofdatatable};
	\addlegendentry{Peak consumption of fast memory with \name}
	\end{axis}
\end{tikzpicture}
}
\caption{Comparison between \textcolor{dong}{peak} memory consumption of DNN models and fast memory size for ResNet variants.}
\label{fig:mem_consumption}
\end{figure}
\end{comment}

\begin{figure}
\centering
\includegraphics[width=0.48\textwidth]{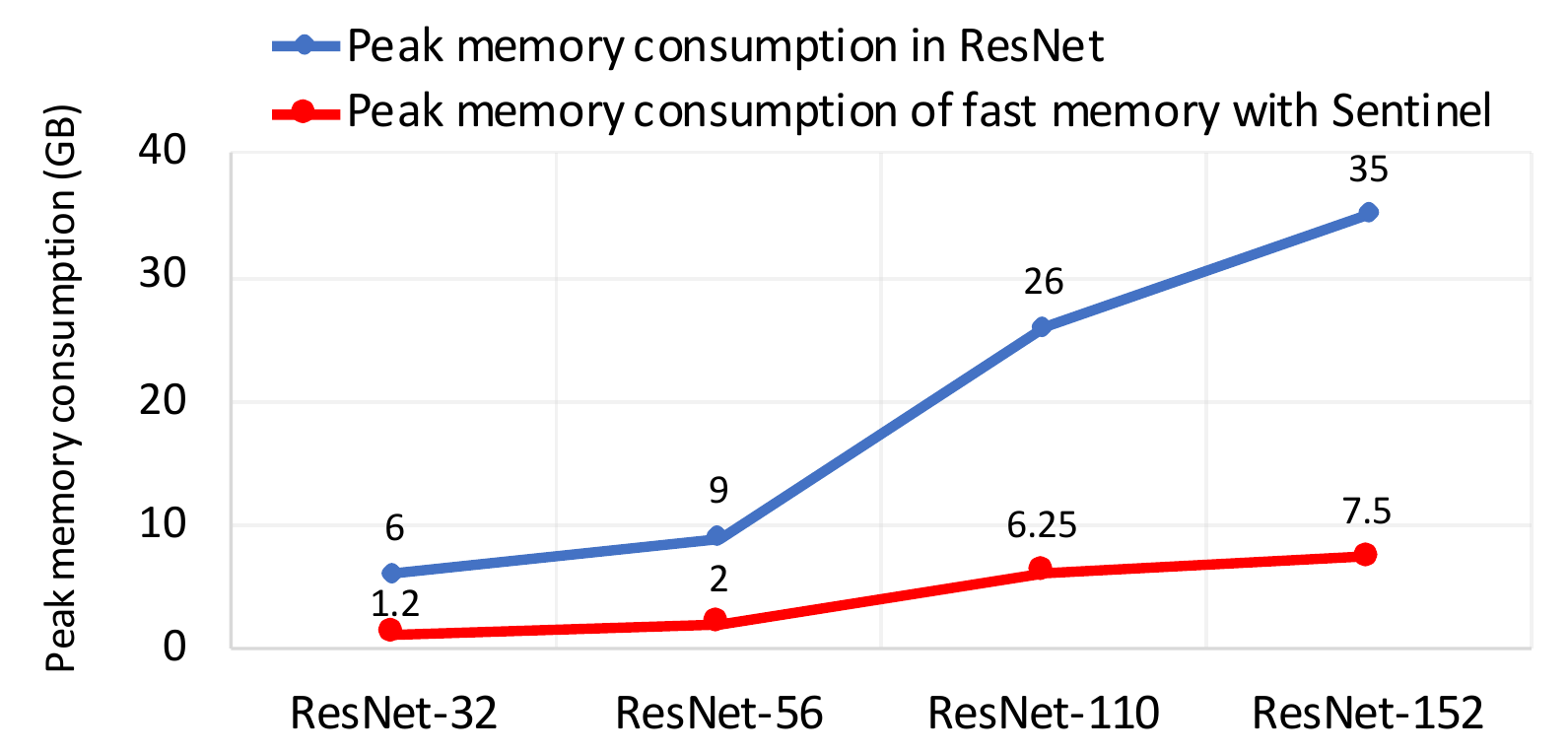}
\caption{Comparison between \textcolor{dong}{peak} memory consumption of DNN models and fast memory size for ResNet variants.}
\label{fig:mem_consumption}
\end{figure}

%\textcolor{red}{Figure: show the cost saving.}
%\textcolor{green}{two bars in one group. five groups in one figure.}

To further study \name's effectiveness, we use various ResNets with various topology. Different ResNets come with different peak memory consumption. 
\textcolor{dong}{We report the minimum fast memory size with which \name performs the same as the fast memory-only.} %\textit{does not have performance loss} with \name, %compared to the fast memory-only cases.
Figure~\ref{fig:mem_consumption} shows peak memory consumption and fast memory size for all ResNet variants. The figure shows that although peak memory consumption increases quickly as ResNet becomes more complicated, the fast memory size increases in a much slower rate. This demonstrates the effectiveness of using \name to save fast memory size.

\begin{comment}
\begin{figure}
\centering
\begin{tikzpicture}
    \begin{axis}[
        width  = 0.5*\textwidth,
        height = 0.3*\textwidth,
        major x tick style = transparent,
        ybar=0.5*\pgflinewidth,
        bar width=10pt,
        %ymajorgrids = true,
        ylabel = {Cost saving},
        symbolic x coords={AA,BB,CC,DD,EE},
        xtick = data,
        scaled y ticks = false,
        enlarge x limits=0.25,
        ymin=0,
        ylabel shift=-0.1cm,
        legend cell align=left,
        legend columns=2,
        legend style={                        at={(1,1.05)},
                anchor=south east,
                column sep=1ex
        }
    ]
        \addplot[style={fill=RYB1,mark=none}]
            coordinates {(AA, 1.0) (BB,1.0) (CC,1.0) (DD,1.0) (EE,1.0)};

        \addplot[style={fill=RYB2,mark=none}]
             coordinates {(AA, 1.0) (BB,1.0) (CC,1.0) (DD,1.0) (EE,1.0)};

        \legend{C1,C2}
    \end{axis}
\end{tikzpicture}
\caption{\textcolor{red}{TODO}}
\end{figure}
\end{comment}
%\textcolor{red}{Figure: number of writes to slow memory for Sentinel and Nimble.}

%% file: text/related_work.tex
\section{Related Work}
\textbf{Heterogeneous main memory.}
%\textcolor{red}{review the existing heterogeneous memory (see the Nimble paper)}
%In today's data centers, vendors use hybrid memory systems rather than single-tier DRAM-based memory systems~\cite{}. 
Many memory technologies~\cite{6509675,3DXPoint_intel,Optane:intelweb,ieeemicro10:lee} have been proposed to build HM. Intel Optane DC persistent memory plus traditional DDR is an example~\cite{8641463,optane:ucsd}; High bandwidth memory (HBM) plus DDR in Intel Knights Landing is another example~\cite{6509675}. Recent research studies data management on HM using hardware-based approaches~\cite{asplos15:agarwal,hetero_mem_arch,qureshi_micro09, ibm_isca09,gpu_pcm_pact13,hpdc16:wu,row_buffer_pcm_iccd12, Ramos:ics11} or OS/software-based approaches~\cite{eurosys16:dulloor,nas16:giardino,asplos16:lin,Narayan:IPDPS2018,Peng:ISMM17,ismm16:shen,sc18:wu,wen:ICS18,unimem:sc17,Yan:ASPLOS19,Yu:ics17}. The common goal of these studies is to achieve a high service rate by leveraging fast memory as much as possible. 
%from fast memory for optimal performance.

\textbf{Page placement policies and mechanisms.}
Existing proposals~\cite{Thermostat:asplos17,RAMinate:socc16,heteros:isca17,unimem:sc17,sc18:wu,Yan:ASPLOS19} explore various page placement polices. They commonly profile memory access to determine page placement. Some work~\cite{Thermostat:asplos17,RAMinate:socc16,heteros:isca17} tracks hot pages by setting and resetting PTE as Sentinel does, but this tracking mechanism can result in very high runtime overhead. To reduce runtime overhead, the existing work commonly limits the amount of pages to profile, which can compromise profiling accuracy. For example, Thermostat~\cite{Thermostat:asplos17} only profiles 0.5\% of total memory pages; If each page is profiled, there could be 4x slowdown. Unimem~\cite{unimem:sc17} and Tahoe~\cite{sc18:wu} use a hardware counter-based approach to periodically count main memory accesses. This method, although being lightweight, can incorrectly count the number of memory accesses for short-lived data objects because of the sampling nature. Unlike the above work, \name leverages domain knowledge, and hence only profiles a small portion of total execution (one training step) without paying large runtime overhead and losing accuracy. Also, \name associates page-level profiling results with data objects, making profiling results more meaningful for data migration. 

%Existing proposals~\cite{Thermostat:asplos17,RAMinate:socc16,heteros:isca17,unimem:sc17,sc18:wu,Yan:ASPLOS19} explore page placement polices to make the best use of a hybrid main memory. These works profile memory access to determine the page placement. Some works~\cite{Thermostat:asplos17,RAMinate:socc16,heteros:isca17} track hot pages by setting and resetting page tables as Sentinel does, but this tracking mechanism can result in very high runtime overhead. To reduce runtime overhead, they must carefully limit the amount of data to profile, which can compromise the accuracy of the profiling. For example, Thermostat~\cite{Thermostat:asplos17} only profiles 0.5\% of the total memory page; otherwise, if each memory page is profiled, 4x runtime overhead is introduced. Unlike the above proposal, \name only profiles one training step due to the repetitive nature of memory access patterns in DNN training. This configuration minimizes the impact of performance profiling on performance. Moreover, the above proposals do not consider application semantics. Given the memory access patterns typically associated with data objects, the lack of application semantics can reduce data placement efficiency and lose the opportunity to improve performance. Leverage customized memory allocation, \name associate page-level profiling result to data objects and determine data migration on data-object level.

Yan et al.~\cite{Yan:ASPLOS19} guides page placement based on an existing Linux page replacement mechanism. Like Linux, this work uses two FIFO queues (active list and inactive list) to make page migration decisions. However, using this design to decide page migration for common short-lived data objects in DNN can be slow and lacks a global view, which wastes valuable fast memory space and causes unnecessary data movement.

%Nimble~\cite{Yan:ASPLOS19} guides page placement based on an existing Linux page replacement mechanism. Like Linux, this work uses two FIFO queues (active list and inactive list) to implemen . Such a design does not work well in DNN training. According to our analysis, there are a large number of short-term objects throughout the training process, but these short-term data objects often have a negligible impact on the performance of DNN training. Tracking and migrating short-term data objects can result in unnecessary costs. Some methods~\cite{sc18:wu,unimem:sc17} also use a hardware counter-based sampling approach to profile memory pages, which may miss the memory accesses of small data objects and result in incorrect page placement decisions.

In terms of page migration mechanism, Yan et al.~\cite{Yan:ASPLOS19} uses multi-threaded migration for single pages and concurrent migration for multiple pages. Bock et al.~\cite{Bock:2014:CPM:2597917.2597924} allow application to execute without waiting for the completion of page migration, by buffering application writes to migrated pages in a hardware buffer. Wang et al.~\cite{7446088} and Seshadri et al.~\cite{Seshadri:2013:RFE:2540708.2540725} enable fast page migration by enhancing DRAM architecture. \name focuses on page migration policy (not mechanism), but the existing efforts are useful to improve performance of \name.

%In terms of page migration mechanism, Nimble~\cite{Yan:ASPLOS19} improves page migration through multi-threaded migration of single pages and concurrent migration of multiple pages. Bock et al.~\cite{Bock:2014:CPM:2597917.2597924} proposes a solution to keep data in the cache to avoid access locks during the page migration. Re-architecting DRAM to limit page migration occurs only in the memory controller without moving data on the chip~\cite{Ryoo:2018:CGA:3205289.3208064,7446088}. Our migration optimization is different from them. We explore a test-and-trial algorithm to adaptively determine the migration interval for optimal performance.

\textbf{Performance optimization for dataflow-based machine learning frameworks.}
Performance optimization for such frameworks attracted a lot of research efforts recently~\cite{222575,pmlr-v80-gao18a,222629,liu:micro18,ipdps19_liu,234946,google:device_placement,google:device_placement_re,222605,DBLP:conf/asplos/SivathanuCSZ19,DBLP:journals/corr/abs-1802-04730,Wang:2018:SDG:3178487.3178491,222611}. 
Some of them~\cite{7472805,Jin:2018:LMR:3274266.3243904,liu:micro18, ipdps19_liu, DBLP:conf/asplos/SivathanuCSZ19,222611} leverage the predictability of DNN workloads to guide operation scheduling and compiler-based performance optimization. Our work also leverages the unique characteristics of DNN, but focuses on data management on HM. 

%Performance optimization for such frameworks attracted a lot of research efforts recently\cite{pmlr-v80-gao18a,liu:micro18,DBLP:journals/corr/abs-1810-08955,google:device_placement,google:device_placement_re, DBLP:conf/asplos/SivathanuCSZ19} studies the performance optimization of a dataflow-based framework. Liu et al.~\cite{liu:micro18} proposes a software/hardware co-design for heterogeneous processing-in-memory systems, which schedules NN training operations on various heterogeneous hardware resources to optimize system energy efficiency. Mirhoseini et al.~\cite{google:device_placement,google:device_placement_re} clusters operations into groups and schedule each group of operations onto different devices. Liu et al.~\cite{DBLP:journals/corr/abs-1810-08955} explores concurrency control that co-runs operations of NN training on many-core platforms to improve hardware utilization.

%Our work is different from the existing efforts on optimizing dataflow-based framework. The existing work studies operation scheduling, while we study data placement on hybrid memory systems to optimize performance.

%% file: text/conclusion.tex
\section{Conclusions}
Runtime data management on HM often uses an application-agnostic approach. It can suffer from high overhead for memory profiling or low accuracy, cause unnecessary data migration, and/or have difficult to hide data migration overhead. In this paper, we use a new angle to examine the data management problem. By introducing limited domain knowledge, we are able to break the fundamental tradeoff between profiling overhead and accuracy, and effectively prefetch data to fast memory for computation. We also reveal the conflict between OS and application when handling data migration. By resolving the conflict, we avoid unnecessary data migration. We focus on a specific and influential domain, DNN, in our study, given its importance on modern data centers. Using \name, DNN training on HM with a small fast memory size can perform similar to the fast memory-only system. Also, \name consistently outperforms a state-of-the-art solution by 18\%.

%Using only 20\% of peak memory consumption of DNN models as the fast memory size, performance of Sentinel is the same or similar (at most 8\% performance difference) to that of the fast memory-only system; 

%Our work shows big performance benefit (\textcolor{red}{xxx}) and reduces the size of valuable fast memory \textcolor{red}{xxx}. 

%% file: ms.bbl
\begin{thebibliography}{10}

\bibitem{dcgan}
{A TensorFlow Implementation of Deep Convolutional Generative Adversarial
  Networks}.
\newblock \url{https://github.com/carpedm20/DCGAN-tensorflow}.

\bibitem{halide}
{{Halide}}.
\newblock https://halide-lang.org/.

\bibitem{intelMKL}
{Intel Math Kernel Library For Deep Neural Networks (Intel MKL-DNN)}.
\newblock \url{https://github.com/intel/mkl-dnn}.

\bibitem{mxnet}
{{MXNet}}.
\newblock https://mxnet.apache.org/.

\bibitem{pytorch}
{{Pytorch}}.
\newblock https://pytorch.org/.

\bibitem{resnet_32}
{ResNet in TensorFlow}.
\newblock \url{https://github.com/wenxinxu/resnet-in-tensorflow}.

\bibitem{tf_models}
{TensorFlow models}.
\newblock \url{https://github.com/tensorflow/models}.

\bibitem{tensorflow2015-whitepaper}
Mart\'{\i}n Abadi, Ashish Agarwal, Paul Barham, Eugene Brevdo, Zhifeng Chen,
  Craig Citro, Greg~S. Corrado, Andy Davis, Jeffrey Dean, Matthieu Devin,
  Sanjay Ghemawat, Ian Goodfellow, Andrew Harp, Geoffrey Irving, Michael Isard,
  Yangqing Jia, Rafal Jozefowicz, Lukasz Kaiser, Manjunath Kudlur, Josh
  Levenberg, Dandelion Man\'{e}, Rajat Monga, Sherry Moore, Derek Murray, Chris
  Olah, Mike Schuster, Jonathon Shlens, Benoit Steiner, Ilya Sutskever, Kunal
  Talwar, Paul Tucker, Vincent Vanhoucke, Vijay Vasudevan, Fernanda Vi\'{e}gas,
  Oriol Vinyals, Pete Warden, Martin Wattenberg, Martin Wicke, Yuan Yu, and
  Xiaoqiang Zheng.
\newblock {TensorFlow}: Large-scale machine learning on heterogeneous systems,
  2015.

\bibitem{asplos15:agarwal}
Neha Agarwal, David Nellans, Mark Stephenson, Mike O'Connor, and Stephen~W.
  Keckler.
\newblock {Page Placement Strategies for GPUs within Heterogeneous Memory
  Systems}.
\newblock In {\em International Conference on Architectural Support for
  Programming Languages and Operating Systems (ASPLOS)}, 2015.

\bibitem{Thermostat:asplos17}
Neha Agarwal and Thomas~F. Wenisch.
\newblock Thermostat: Application-transparent page management for two-tiered
  main memory.
\newblock In {\em Proceedings of the Twenty-Second International Conference on
  Architectural Support for Programming Languages and Operating Systems,
  {ASPLOS} 2017, Xi'an, China, April 8-12, 2017}, pages 631--644, 2017.

\bibitem{8641463}
M.~{Arafa}, B.~{Fahim}, S.~{Kottapalli}, A.~{Kumar}, L.~P. {Looi},
  S.~{Mandava}, A.~{Rudoff}, I.~M. {Steiner}, B.~{Valentine}, G.~{Vedaraman},
  and S.~{Vora}.
\newblock Cascade lake: Next generation intel xeon scalable processor.
\newblock {\em IEEE Micro}, 39(2):29--36, March 2019.

\bibitem{arm_cl}
ARM.
\newblock {{ARM Compute Library}}.
\newblock https://github.com/ARM-software/ComputeLibrary.

\bibitem{DBLP:journals/corr/abs-1901-10938}
Joy Arulraj, Andy Pavlo, and Krishna~Teja Malladi.
\newblock Multi-tier buffer management and storage system design for
  non-volatile memory.
\newblock {\em CoRR}, abs/1901.10938, 2019.

\bibitem{hetero_mem_arch}
Alan Bivens, Parijat Dube, Michele Franceschini, John Karidis, Luis Lastras,
  and Mickey Tsao.
\newblock Architectural design for next generation heterogeneous memory
  systems.
\newblock In {\em 2010 {IEEE} Int. Memory Workshop}, pages 1--4, 2010.

\bibitem{Bock:2014:CPM:2597917.2597924}
Santiago Bock, Bruce~R. Childers, Rami Melhem, and Daniel Moss{\'e}.
\newblock Concurrent page migration for mobile systems with os-managed hybrid
  memory.
\newblock In {\em Proceedings of the 11th ACM Conference on Computing
  Frontiers}, CF '14, pages 31:1--31:10, New York, NY, USA, 2014. ACM.

\bibitem{tacc_stampede}
Texas Advanced~Computing Center.
\newblock {{Stampede2}}.
\newblock https://www.tacc.utexas.edu/systems/stampede26cf c665cv 5.

\bibitem{6509675}
D.~W. {Chang}, G.~{Byun}, H.~{Kim}, M.~{Ahn}, S.~{Ryu}, N.~S. {Kim}, and
  M.~{Schulte}.
\newblock Reevaluating the latency claims of 3d stacked memories.
\newblock In {\em 2013 18th Asia and South Pacific Design Automation Conference
  (ASP-DAC)}, pages 657--662, Jan 2013.

\bibitem{7472805}
K.~{Chen} and Q.~{Huo}.
\newblock Scalable training of deep learning machines by incremental block
  training with intra-block parallel optimization and blockwise model-update
  filtering.
\newblock In {\em 2016 IEEE International Conference on Acoustics, Speech and
  Signal Processing (ICASSP)}, pages 5880--5884, March 2016.

\bibitem{222575}
Tianqi Chen, Thierry Moreau, Ziheng Jiang, Lianmin Zheng, Eddie Yan, Haichen
  Shen, Meghan Cowan, Leyuan Wang, Yuwei Hu, Luis Ceze, Carlos Guestrin, and
  Arvind Krishnamurthy.
\newblock {TVM}: An automated end-to-end optimizing compiler for deep learning.
\newblock In {\em 13th {USENIX} Symposium on Operating Systems Design and
  Implementation ({OSDI} 18)}, pages 578--594, Carlsbad, CA, October 2018.
  {USENIX} Association.

\bibitem{7738524}
Y.~{Chen}, T.~{Krishna}, J.~S. {Emer}, and V.~{Sze}.
\newblock {Eyeriss: An Energy-Efficient Reconfigurable Accelerator for Deep
  Convolutional Neural Networks}.
\newblock {\em IEEE Journal of Solid-State Circuits}, 52(1):127--138, 2017.

\bibitem{DBLP:journals/corr/ChengKHSCAACCIA16}
Heng{-}Tze Cheng, Levent Koc, Jeremiah Harmsen, Tal Shaked, Tushar Chandra,
  Hrishi Aradhye, Glen Anderson, Greg Corrado, Wei Chai, Mustafa Ispir, Rohan
  Anil, Zakaria Haque, Lichan Hong, Vihan Jain, Xiaobing Liu, and Hemal Shah.
\newblock {Wide {\&} Deep Learning for Recommender Systems}.
\newblock {\em CoRR}, abs/1606.07792, 2016.

\bibitem{DeVito:2011:LDS:2063384.2063396}
Zachary DeVito, Niels Joubert, Francisco Palacios, Stephen Oakley, Montserrat
  Medina, Mike Barrientos, Erich Elsen, Frank Ham, Alex Aiken, Karthik
  Duraisamy, Eric Darve, Juan Alonso, and Pat Hanrahan.
\newblock {Liszt: A Domain Specific Language for Building Portable Mesh-based
  PDE Solvers}.
\newblock In {\em International Conference for High Performance Computing,
  Networking, Storage and Analysis}, 2011.

\bibitem{eurosys16:dulloor}
Subramanya~R. Dulloor, Amitabha Roy, Zheguang Zhao, Narayanan Sundaram,
  Nadathur Satish, Rajesh Sankaran, Jeff Jackson, and Karsten Schwan.
\newblock Data tiering in heterogeneous memory systems.
\newblock In {\em Proc. 11th European Conf. Computer Systems ({EuroSys} '16)},
  2016.

\bibitem{Eisenman:2018:RDF:3190508.3190524}
Assaf Eisenman, Darryl Gardner, Islam AbdelRahman, Jens Axboe, Siying Dong, Kim
  Hazelwood, Chris Petersen, Asaf Cidon, and Sachin Katti.
\newblock {Reducing DRAM Footprint with NVM in Facebook}.
\newblock In {\em Proceedings of the Thirteenth EuroSys Conference}, 2018.

\bibitem{pmlr-v80-gao18a}
Yuanxiang Gao, Li~Chen, and Baochun Li.
\newblock Spotlight: Optimizing device placement for training deep neural
  networks.
\newblock In Jennifer Dy and Andreas Krause, editors, {\em Proceedings of the
  35th International Conference on Machine Learning}, volume~80 of {\em
  Proceedings of Machine Learning Research}, pages 1676--1684,
  Stockholmsmässan, Stockholm Sweden, 10--15 Jul 2018. PMLR.

\bibitem{nas16:giardino}
Michael Giardino, Kshitij Doshi, and Bonnie Ferri.
\newblock {Soft2LM}: Application guided heterogeneous memory management.
\newblock In {\em 2016 {IEEE} Int. Conf. Networking, Architecture, and Storage
  (NAS)}, 2016.

\bibitem{google_tensorflow_bucketing}
{{Google}}.
\newblock {Tensorflow Bucketing}.
\newblock https:// www.tensorflow.org/versions/r0.12/.

\bibitem{3DXPoint_intel}
F.~T. {Hady}, A.~{Foong}, B.~{Veal}, and D.~{Williams}.
\newblock Platform storage performance with 3d xpoint technology.
\newblock {\em Proceedings of the IEEE}, 105(9):1822--1833, Sep. 2017.

\bibitem{Hasselt:2016:DRL:3016100.3016191}
Hado~van Hasselt, Arthur Guez, and David Silver.
\newblock {Deep Reinforcement Learning with Double Q-Learning}.
\newblock In {\em Proceedings of the Thirtieth AAAI Conference on Artificial
  Intelligence}, 2016.

\bibitem{RAMinate:socc16}
Takahiro Hirofuchi and Ryousei Takano.
\newblock Raminate: Hypervisor-based virtualization for hybrid main memory
  systems.
\newblock In {\em Proceedings of the Seventh ACM Symposium on Cloud Computing},
  SoCC '16, pages 112--125, New York, NY, USA, 2016. ACM.

\bibitem{Optane:intelweb}
Intel.
\newblock {Revolutionizing Memory and Storage}.
\newblock
  \url{https://www.intel.com/content/www/us/en/architecture-and-technology/intel-optane-technology.html}.

\bibitem{optane:ucsd}
Joseph Izraelevitz, Jian Yang, Lu~Zhang, Juno Kim, Xiao Liu, Amirsaman
  Memaripour, Yun~Joon Soh, Zixuan Wang, Yi~Xu, Subramanya~R. Dulloor, Jishen
  Zhao, and Steven Swanson.
\newblock Basic performance measurements of the intel optane {DC} persistent
  memory module.
\newblock {\em CoRR}, abs/1903.05714, 2019.

\bibitem{hbm}
JEDEC.
\newblock {{JESD79-4A: DDR4 SDRAM Standard}}.
\newblock https://www. jedec.org/sites/default/files/docs/JESD79-4A.pdf.

\bibitem{Jin:2018:LMR:3274266.3243904}
Hai Jin, Bo~Liu, Wenbin Jiang, Yang Ma, Xuanhua Shi, Bingsheng He, and Shaofeng
  Zhao.
\newblock Layer-centric memory reuse and data migration for extreme-scale deep
  learning on many-core architectures.
\newblock {\em ACM Trans. Archit. Code Optim.}, 15(3):37:1--37:26, September
  2018.

\bibitem{Jouppi:2017:IPA:3079856.3080246}
Norman~P. Jouppi, Cliff Young, Nishant Patil, David Patterson, Gaurav Agrawal,
  Raminder Bajwa, Sarah Bates, Suresh Bhatia, Nan Boden, Al~Borchers, Rick
  Boyle, Pierre-luc Cantin, Clifford Chao, Chris Clark, Jeremy Coriell, Mike
  Daley, Matt Dau, Jeffrey Dean, Ben Gelb, Tara~Vazir Ghaemmaghami, Rajendra
  Gottipati, William Gulland, Robert Hagmann, C.~Richard Ho, Doug Hogberg, John
  Hu, Robert Hundt, Dan Hurt, Julian Ibarz, Aaron Jaffey, Alek Jaworski,
  Alexander Kaplan, Harshit Khaitan, Daniel Killebrew, Andy Koch, Naveen Kumar,
  Steve Lacy, James Laudon, James Law, Diemthu Le, Chris Leary, Zhuyuan Liu,
  Kyle Lucke, Alan Lundin, Gordon MacKean, Adriana Maggiore, Maire Mahony,
  Kieran Miller, Rahul Nagarajan, Ravi Narayanaswami, Ray Ni, Kathy Nix, Thomas
  Norrie, Mark Omernick, Narayana Penukonda, Andy Phelps, Jonathan Ross, Matt
  Ross, Amir Salek, Emad Samadiani, Chris Severn, Gregory Sizikov, Matthew
  Snelham, Jed Souter, Dan Steinberg, Andy Swing, Mercedes Tan, Gregory
  Thorson, Bo~Tian, Horia Toma, Erick Tuttle, Vijay Vasudevan, Richard Walter,
  Walter Wang, Eric Wilcox, and Doe~Hyun Yoon.
\newblock {In-Datacenter Performance Analysis of a Tensor Processing Unit}.
\newblock In {\em International Symposium on Computer Architecture}, 2017.

\bibitem{heteros:isca17}
S.~{Kannan}, A.~{Gavrilovska}, V.~{Gupta}, and K.~{Schwan}.
\newblock Heteroos — os design for heterogeneous memory management in
  datacenter.
\newblock In {\em 2017 ACM/IEEE 44th Annual International Symposium on Computer
  Architecture (ISCA)}, pages 521--534, June 2017.

\bibitem{8658402}
S.~{Kim}, H.~{Kim}, J.~{Lee}, S.~{Yoon}, S.~E. {Kahou}, K.~{Kashinath}, and
  M.~{Prabhat}.
\newblock {Deep-Hurricane-Tracker: Tracking and Forecasting Extreme Climate
  Events}.
\newblock In {\em IEEE Winter Conference on Applications of Computer Vision
  (WACV)}, 2019.

\bibitem{Kurth:2018:EDL:3291656.3291724}
Thorsten Kurth, Sean Treichler, Joshua Romero, Mayur Mudigonda, Nathan Luehr,
  Everett Phillips, Ankur Mahesh, Michael Matheson, Jack Deslippe, Massimiliano
  Fatica, Prabhat, and Michael Houston.
\newblock {Exascale Deep Learning for Climate Analytics}.
\newblock In {\em Proceedings of the International Conference for High
  Performance Computing, Networking, Storage, and Analysis}, 2018.

\bibitem{lbnl_cori}
Lawerence Berkeley~National Lab.
\newblock {{Cori Supercomputer}}.
\newblock https://www.nersc.gov/users/computational-systems/cori/.

\bibitem{ieeemicro10:lee}
Benjamin~C. Lee, Ping Zhou, Jun Yang, Youtao Zhang, Bo~Zhao, Engin Ipek, Onur
  Mutlu, and Doug Burger.
\newblock {Phase-Change Technology and the Future of Main Memory}.
\newblock {\em IEEE Micro}, 30(1):143--143, 2010.

\bibitem{222629}
Yunseong Lee, Alberto Scolari, Byung-Gon Chun, Marco~Domenico Santambrogio,
  Markus Weimer, and Matteo Interlandi.
\newblock {PRETZEL}: Opening the black box of machine learning prediction
  serving systems.
\newblock In {\em 13th {USENIX} Symposium on Operating Systems Design and
  Implementation ({OSDI} 18)}, pages 611--626, Carlsbad, CA, October 2018.
  {USENIX} Association.

\bibitem{asplos16:lin}
Felix~Xiaozhu Lin and Xu~Liu.
\newblock {memif: Towards Programming Heterogeneous Memory Asynchronously}.
\newblock In {\em International Conference on Architectural Support for
  Programming Languages and Operating Systems (ASPLOS)}, 2016.

\bibitem{liu:micro18}
J.~{Liu}, H.~{Zhao}, M.~A. {Ogleari}, D.~{Li}, and J.~{Zhao}.
\newblock Processing-in-memory for energy-efficient neural network training: A
  heterogeneous approach.
\newblock In {\em 2018 51st Annual IEEE/ACM International Symposium on
  Microarchitecture (MICRO)}, pages 655--668, Oct 2018.

\bibitem{ipdps19_liu}
Jiawen Liu, Dong Li, Gokcen Kestor, and Jeffrey~S. Vetter.
\newblock {Runtime Concurrency Control and Operation Scheduling for High
  Performance Neural Network Training}.
\newblock In {\em International Parallel and Distributed Processing Symposium},
  2019.

\bibitem{234946}
Yizhi Liu, Yao Wang, Ruofei Yu, Mu~Li, Vin Sharma, and Yida Wang.
\newblock Optimizing {CNN} model inference on cpus.
\newblock In {\em 2019 {USENIX} Annual Technical Conference ({USENIX} {ATC}
  19)}, pages 1025--1040, Renton, WA, July 2019. {USENIX} Association.

\bibitem{ram_price}
Jon Martindale.
\newblock {{RAM has never been cheaper, but are the historic prices here to
  stay? }}.
\newblock https://www.digitaltrends.com/computing/why-is-ram-so-cheap/.

\bibitem{Mathuriya:2018:CUD:3291656.3291743}
Amrita Mathuriya, Deborah Bard, Peter Mendygral, Lawrence Meadows, James
  Arnemann, Lei Shao, Siyu He, Tuomas K\"{a}rn\"{a}, Diana Moise, Simon~J.
  Pennycook, Kristyn Maschhoff, Jason Sewall, Nalini Kumar, Shirley Ho,
  Michael~F. Ringenburg, Prabhat, and Victor Lee.
\newblock {CosmoFlow: Using Deep Learning to Learn the Universe at Scale}.
\newblock In {\em Proceedings of the International Conference for High
  Performance Computing, Networking, Storage, and Analysis}, 2018.

\bibitem{google:device_placement}
Azalia Mirhoseini, Anna Goldie, Hieu Pham, Benoit Steiner, Quoc~V. Le, and Jeff
  Dean.
\newblock Hierarchical planning for device placement.
\newblock 2018.

\bibitem{google:device_placement_re}
Azalia Mirhoseini, Hieu Pham, Quoc Le, Mohammad Norouzi, Samy Bengio, Benoit
  Steiner, Yuefeng Zhou, Naveen Kumar, Rasmus Larsen, and Jeff Dean.
\newblock Device placement optimization with reinforcement learning.
\newblock 2017.

\bibitem{222605}
Philipp Moritz, Robert Nishihara, Stephanie Wang, Alexey Tumanov, Richard Liaw,
  Eric Liang, Melih Elibol, Zongheng Yang, William Paul, Michael~I. Jordan, and
  Ion Stoica.
\newblock Ray: A distributed framework for emerging {AI} applications.
\newblock In {\em 13th {USENIX} Symposium on Operating Systems Design and
  Implementation ({OSDI} 18)}, pages 561--577, Carlsbad, CA, October 2018.
  {USENIX} Association.

\bibitem{Narayan:IPDPS2018}
A.~{Narayan}, T.~{Zhang}, S.~{Aga}, S.~{Narayanasamy}, and A.~{Coskun}.
\newblock Moca: Memory object classification and allocation in heterogeneous
  memory systems.
\newblock In {\em 2018 IEEE International Parallel and Distributed Processing
  Symposium (IPDPS)}, 2018.

\bibitem{8192478}
A.~{Parashar}, M.~{Rhu}, A.~{Mukkara}, A.~{Puglielli}, R.~{Venkatesan},
  B.~{Khailany}, J.~{Emer}, S.~W. {Keckler}, and W.~J. {Dally}.
\newblock {SCNN: An accelerator for compressed-sparse convolutional neural
  networks}.
\newblock In {\em International Symposium on Computer Architecture (ISCA)},
  2017.

\bibitem{Peng:ISMM17}
Ivy~Bo Peng, Roberto Gioiosa, Gokcen Kestor, Pietro Cicotti, Erwin Laure, and
  Stefano Markidis.
\newblock Rthms: A tool for data placement on hybrid memory system.
\newblock In {\em Proceedings of the 2017 ACM 9t9999999 International Symposium
  on Memory Management}, ISMM 2017, 2017.

\bibitem{democratization_ai}
Tom Petrocelli.
\newblock {{Democratization of AI Development Is Beginning}}.
\newblock
  https://www.cmswire.com/digital-workplace/democratization-of-ai-development-is-beginning/.

\bibitem{qureshi_micro09}
Moinuddin~K. Qureshi, Michele Franchescini, Vijayalakshmi Srinivasan, Luis
  Lastras, Bulent Abali, and John Karidis.
\newblock {Enhancing Lifetime and Security of PCM-Based Main Memory with
  Start-Gap Wear Leveling}.
\newblock In {\em MICRO}, 2009.

\bibitem{ibm_isca09}
Moinuddin~K. Qureshi, Viji Srinivasan, and Jude~A. Rivers.
\newblock {Scalable High-Performance Main Memory System Using Phase-Change
  Memory Technology}.
\newblock In {\em ISCA}, 2009.

\bibitem{Ramos:ics11}
Luiz Ramos, Eugene Gorbatov, and Ricardo Bianchini.
\newblock Page placement in hybrid memory systems.
\newblock In {\em Proc. Int. Conf. Supercomputing ({ICS} '11)}, 2011.

\bibitem{ram_price2}
Gina Roos.
\newblock {{Dram Prices Continue to Climb}}.
\newblock https://epsnews.com/2017/08/18/dram-prices-continue-climb/.

\bibitem{Seshadri:2013:RFE:2540708.2540725}
Vivek Seshadri, Yoongu Kim, Chris Fallin, Donghyuk Lee, Rachata
  Ausavarungnirun, Gennady Pekhimenko, Yixin Luo, Onur Mutlu, Phillip~B.
  Gibbons, Michael~A. Kozuch, and Todd~C. Mowry.
\newblock {RowClone: Fast and Energy-efficient in-DRAM Bulk Data Copy and
  Initialization}.
\newblock In {\em IEEE/ACM International Symposium on Microarchitecture}, 2013.

\bibitem{Shafiee:2016:ICN:3001136.3001139}
Ali Shafiee, Anirban Nag, Naveen Muralimanohar, Rajeev Balasubramonian,
  John~Paul Strachan, Miao Hu, R.~Stanley Williams, and Vivek Srikumar.
\newblock {ISAAC: A Convolutional Neural Network Accelerator with In-situ
  Analog Arithmetic in Crossbars}.
\newblock In {\em International Symposium on Computer Architecture}, 2016.

\bibitem{Shaw:2008:ASM:1364782.1364802}
David~E. Shaw, Martin~M. Deneroff, Ron~O. Dror, Jeffrey~S. Kuskin, Richard~H.
  Larson, John~K. Salmon, Cliff Young, Brannon Batson, Kevin~J. Bowers, Jack~C.
  Chao, Michael~P. Eastwood, Joseph Gagliardo, J.~P. Grossman, C.~Richard Ho,
  Douglas~J. Ierardi, Istv\'{a}n Kolossv\'{a}ry, John~L. Klepeis, Timothy
  Layman, Christine McLeavey, Mark~A. Moraes, Rolf Mueller, Edward~C. Priest,
  Yibing Shan, Jochen Spengler, Michael Theobald, Brian Towles, and Stanley~C.
  Wang.
\newblock {Anton, a Special-purpose Machine for Molecular Dynamics Simulation}.
\newblock {\em Communications of the ACM}, 51(7), July 2008.

\bibitem{ismm16:shen}
Du~Shen, Xu~Liu, and Felix~Xiaozhu Lin.
\newblock {Characterizing Emerging Heterogeneous Memory}.
\newblock In {\em ACM SIGPLAN International Symposium on Memory Management
  (ISMM)}, 2016.

\bibitem{DBLP:conf/asplos/SivathanuCSZ19}
Muthian Sivathanu, Tapan Chugh, Sanjay~S. Singapuram, and Lidong Zhou.
\newblock {Astra: Exploiting Predictability to Optimize Deep Learning}.
\newblock In {\em International Conference on Architectural Support for
  Programming Languages and Operating Systems}, 2019.

\bibitem{tacc_ml_CPU_training}
Vamsi Sripathi and Vikram Saletore.
\newblock {{TensorFlow Performance Optimization on Intel Architecture}}.
\newblock https://www.alcf.anl.gov/files/slides \%20vamsi \%20sripathi
  \%20vikram \%20saletore \%20ACLF\_ANL\_07253018\_final.pdf.

\bibitem{DBLP:journals/corr/abs-1802-04730}
Nicolas Vasilache, Oleksandr Zinenko, Theodoros Theodoridis, Priya Goyal,
  Zachary DeVito, William~S. Moses, Sven Verdoolaege, Andrew Adams, and Albert
  Cohen.
\newblock Tensor comprehensions: Framework-agnostic high-performance machine
  learning abstractions.
\newblock {\em CoRR}, abs/1802.04730, 2018.

\bibitem{gpu_pcm_pact13}
Bin Wang, Bo~Wu, Dong Li, Xipeng Shen, Weikuan Yu, Yizheng Jiao, and Jeffrey~S.
  Vetter.
\newblock Exploring hybrid memory for {GPU} energy efficiency through
  software-hardware co-design.
\newblock In {\em Proc. 22nd Int. Conf. Parallel Architectures and Compilation
  Techniques ({PACT} '13)}, 2013.

\bibitem{7446088}
H.~{Wang}, J.~{Zhang}, S.~{Shridhar}, G.~{Park}, M.~{Jung}, and N.~S. {Kim}.
\newblock Duang: Fast and lightweight page migration in asymmetric memory
  systems.
\newblock In {\em 2016 IEEE International Symposium on High Performance
  Computer Architecture (HPCA)}, pages 481--493, March 2016.

\bibitem{Wang:2018:SDG:3178487.3178491}
Linnan Wang, Jinmian Ye, Yiyang Zhao, Wei Wu, Ang Li, Shuaiwen~Leon Song,
  Zenglin Xu, and Tim Kraska.
\newblock Superneurons: Dynamic gpu memory management for training deep neural
  networks.
\newblock In {\em Proceedings of the 23rd ACM SIGPLAN Symposium on Principles
  and Practice of Parallel Programming}, PPoPP '18, pages 41--53, New York, NY,
  USA, 2018. ACM.

\bibitem{wen:ICS18}
Shasha Wen, Lucy Cherkasova, Felix~Xiaozhu Lin, and Xu~Liu.
\newblock Profdp: A lightweight profiler to guide data placement in
  heterogeneous memory systems.
\newblock In {\em Proceedings of the 2018 International Conference on
  Supercomputing}, ICS '18, 2018.

\bibitem{sc18:wu}
K.~Wu, J.~Ren, and D.~Li.
\newblock Runtime data management on non-volatile memory-based heterogeneous
  memory for task-parallel programs.
\newblock In {\em Proceedings of the International Conference for High
  Performance Computing, Networking, Storage, and Analysis}, 2018.

\bibitem{unimem:sc17}
Kai Wu, Yingchao Huang, and Dong Li.
\newblock {Unimem: Runtime Data Management on Non-volatile Memory-based
  Heterogeneous Main Memory}.
\newblock In {\em SC}, 2017.

\bibitem{hpdc16:wu}
Panruo Wu, Dong Li, Zizhong Chen, Jeffrey Vetter, and Sparsh Mittal.
\newblock {Algorithm-Directed Data Placement in Explicitly Managed No-Volatile
  Memory}.
\newblock In {\em Proc. 25th ACM Int. Symp. High-Performance Parallel and
  Distributed Computing}, pages 141--152, Kyoto, Japan, 2016.

\bibitem{222611}
Wencong Xiao, Romil Bhardwaj, Ramachandran Ramjee, Muthian Sivathanu, Nipun
  Kwatra, Zhenhua Han, Pratyush Patel, Xuan Peng, Hanyu Zhao, Quanlu Zhang, Fan
  Yang, and Lidong Zhou.
\newblock Gandiva: Introspective cluster scheduling for deep learning.
\newblock In {\em 13th {USENIX} Symposium on Operating Systems Design and
  Implementation ({OSDI} 18)}, pages 595--610, Carlsbad, CA, October 2018.
  {USENIX} Association.

\bibitem{Yan:ASPLOS19}
Zi~Yan, Daniel Lustig, David Nellans, and Abhishek Bhattacharjee.
\newblock Nimble page management for tiered memory systems.
\newblock In {\em Proceedings of the Twenty-Fourth International Conference on
  Architectural Support for Programming Languages and Operating Systems},
  ASPLOS '19, 2019.

\bibitem{row_buffer_pcm_iccd12}
HanBin Yoon, Justin Meza, Rachata Ausavarungnirun, Rachael Harding, and Onur
  Mutlu.
\newblock Row buffer locality aware caching policies for hybrid memories.
\newblock In {\em Proc. {IEEE} 2012 30th Int. Conf. Computer Design ({ICCD}
  '12)}, 2012.

\bibitem{Yu:ics17}
Seongdae Yu, Seongbeom Park, and Woongki Baek.
\newblock {Design and Implementation of Bandwidth-aware Memory Placement and
  Migration Policies for Heterogeneous Memory Systems}.
\newblock In {\em Proceedings of the International Conference on Supercomputing
  (ICS)}, 2017.

\bibitem{5260554}
W.~{Zhang} and T.~{Li}.
\newblock {Exploring Phase Change Memory and 3D Die-Stacking for Power/Thermal
  Friendly, Fast and Durable Memory Architectures}.
\newblock In {\em International Conference on Parallel Architectures and
  Compilation Techniques (PACT)}, 2009.

\end{thebibliography}
